\documentclass[prx,twocolumn,preprintnumbers,amsmath,amssymb,superscriptaddress,longbibliography]{revtex4-2}

\usepackage{color}
\usepackage{graphicx}
\usepackage{dcolumn}
\usepackage{bm}
\usepackage{hyperref}
\usepackage{enumitem}
\usepackage{balance}
\usepackage{comment}
\usepackage{cleveref}

\usepackage{amsthm, braket}
\usepackage[normalem]{ulem}

\begin{document}
\title{Ultrahigh threshold nonstabilizer nonlinear quantum error correcting code}

\author{Maga Grafe}
\thanks{These authors contributed equally}
\affiliation{New York University Shanghai, NYU-ECNU Institute of Physics at NYU Shanghai, 567 West Yangsi Road, Shanghai, 200124, China}
\affiliation{PSL Dauphine University, Place du Maréchal de Lattre de Tassigny, 75016 Paris, France}

\author{Kaixuan Zhou}
\thanks{These authors contributed equally}
\affiliation{New York University Shanghai, NYU-ECNU Institute of Physics at NYU Shanghai, 567 West Yangsi Road, Shanghai, 200124, China}
\affiliation{Department of Physics, New York University, New York, NY 10003, USA}

\author{Zaman Tekin}
\affiliation{New York University Shanghai, NYU-ECNU Institute of Physics at NYU Shanghai, 567 West Yangsi Road, Shanghai, 200124, China}
\affiliation{Mathematical Institute, University of Oxford, Radcliffe Observatory, Andrew Wiles Building, Woodstock Rd, Oxford OX2 6GG, United Kingdom}

\author{Zhiyuan Lin}
\affiliation{State Key Laboratory of Precision Spectroscopy, School of Physical and Material Sciences, East China Normal University, Shanghai 200062, China}

\author{Sen Li}
\affiliation{State Key Laboratory of Precision Spectroscopy, School of Physical and Material Sciences, East China Normal University, Shanghai 200062, China}

\author{Fengquan Zhang}
\affiliation{State Key Laboratory of Precision Spectroscopy, School of Physical and Material Sciences, East China Normal University, Shanghai 200062, China}

\author{Valentin Ivannikov}
\affiliation{New York University Shanghai, NYU-ECNU Institute of Physics at NYU Shanghai, 567 West Yangsi Road, Shanghai, 200124, China}

\author{Tim Byrnes}
 \email{tim.byrnes@nyu.edu}
\affiliation{New York University Shanghai, NYU-ECNU Institute of Physics at NYU Shanghai, 567 West Yangsi Road, Shanghai, 200124, China}
\affiliation{State Key Laboratory of Precision Spectroscopy, School of Physical and Material Sciences, East China Normal University, Shanghai 200062, China}
\affiliation{Center for Quantum and Topological Systems (CQTS), NYUAD Research Institute, New York University Abu Dhabi, UAE.}
\affiliation{Department of Physics, New York University, New York, NY 10003, USA}

\begin{abstract}
We introduce a novel type of quantum error correcting code, called the spinor code, based on spaces defined by total spin.  The code is a nonstabilizer code, and is also a nonlinear quantum error correcting code, meaning that quantum information is encoded in a parameterized family of quantum states, rather than a linear superposition of code words.  
Syndrome measurements are performed by projecting on states with differing total spin, with an associated correction to map states back to the maximum total spin space.  We show that the code is asymptotically capable of protecting against any single qubit Pauli error for Gaussian distributed states such as spin coherent state. We directly evaluate the performance under the depolarizing channel, considering various cases, with and without initialization and measurement errors, as well as two qubit errors.  We estimate the code-capacity threshold to be in the range of 32-75\%, while the phenomenological threshold is in the range 9-75\%.   
\end{abstract}

\date{\today}

\maketitle

\section{Introduction}

It was realized early on that quantum error correction (QEC) \cite{shor1995scheme,gottesman1997stabilizer,peres1985reversible} is an essential component for realizing reliable large-scale quantum computers, due to the sensitivity of quantum states to decoherence \cite{devitt2013quantum,roffe2019quantum,terhal2015quantum,steane2006tutorial,nielsen2002quantum}. {
The basic strategy in a QEC code is to define logical code words which involve some level of redundancy, such that it is possible to perform a quantum measurement that identifies the type of error, then perform a correction operation to revert the state to its original form.  The type of structure that is required for a QEC code is encapsulated by the Knill-Laflamme criterion \cite{knill1997theory}, which gives necessary and sufficient conditions for a correctable code.  In short, one requires that the various error spaces are orthogonal, such that different types of errors can be distinguished; furthermore, the quantum information should not be corrupted (deformed) during the measurement-correction step. One of the key results of quantum fault-tolerance is the threshold theorem \cite{shor1996fault,aharonov1997fault,knill1998resilient,kitaev2003fault}, which states that by scaling up a QEC, it is possible to suppress logical errors to arbitrarily low levels as long as the physical error rate is below a threshold value.  Numerous types of QEC codes have been developed to meet the needs of large-scale quantum computers to date.   Early examples include Shor's repetition code \cite{shor1995scheme}, Steane's code\cite{steane1996multiple}, and  Calderbank–Shor–Steane (CSS) codes \cite{calderbank1996good,steane1996simple}.  More recently, the surface code \cite{raussendorf2007fault,fowler2012surface} is one of the most-studied codes, and is a variation of the toric code that was originally introduced by Kitaev \cite{kitaev2003fault,dennis2002topological}.  Its popularity stems from high error thresholds \cite{dennis2002topological,wang2011threshold,bombin2012strong,wootton2012high, fowler2012surface, tuckett2018ultrahigh}, as well as a two dimensional architecture which makes it compatible with existing quantum computer architectures. Several other state-of-the-art codes also give competitive error thresholds such as the quantum low density parity check (qLDPC) code \cite{breuckmann2021quantum,bravyi2024high},  fusion based quantum computing \cite{bartolucci2023fusion}, XZZX surface code \cite{bonilla2021xzzx,wu2022erasure}, and 
subsystem surface codes \cite{bravyi2012subsystem,higgott2021subsystem}. 
Great strides in experimental progress of demonstrating QEC have been made in recent years  \cite{cory1998experimental,pittman2005demonstration,chiaverini2004realization,schindler2011experimental,reed2012realization,ofek2016extending,ryan2021realization,krinner2022realizing,livingston2022experimental,sivak2023real,google2023suppressing,bluvstein2024logical,gupta2024encoding,paetznick2024demonstration,acharya2025quantum}.  

The majority of QEC codes that have been developed to date are stabilizer codes.  A stabilizer code has a code space 
defined by the simultaneous eigenspace of a commuting set of Pauli strings.  The logical code words $ |0_L \rangle, |1_L \rangle $ can then be chosen to be stabilizer states.  One of the first examples of a non-stabilizer (or non-additive) code which uses 5 qubits to encode 6 logical code words \cite{rains1997nonadditive}.  Other qubit-based non-stabilizer codes include the codeword stabilized (CWS) codes \cite{cross2008codeword,chuang2009codeword}, union stabilized codes \cite{grassl2009generalized}, Smolin-Smith-Wehner codes \cite{smolin2007simple}, and Ruskai-Ouyang permutation invariant codes \cite{ruskai1999pauli,pollatsek2004permutationally,ouyang2014permutation,ouyang2026theory}.  In particular, permutation invariant codes are based on using Dicke states to construct logical codewords, and have been extended to qudits \cite{ouyang2017permutation}, constant-excitation encoding \cite{ouyang2019permutation}, other variants \cite{aydin2024family,aydin2026quantum,ouyang2025measurement,kubischta2023family}.  Other examples of non-stabilizer codes are those based on bosonic modes, such as cat codes \cite{mirrahimi2014dynamically}, binomial codes \cite{michael2016new}, and Gottesman-Kitaev-Preskill (GKP) codes \cite{PhysRevA.64.012310}.  Furthermore, to our knowledge virtually all QEC codes are linear codes.  This means that a qubit state is encoded as
\begin{align}
\alpha |0 \rangle + \beta |1 \rangle \rightarrow \alpha |0_L \rangle + \beta |1_L \rangle .
\end{align}
In Ref. \cite{reichert2022nonlinear} the concept of a nonlinear QEC was introduced, where the encoding consists of a parameterized family of quantum states
\begin{align}
\alpha |0 \rangle + \beta |1 \rangle \rightarrow  |\psi(\alpha,\beta) \rangle .
\end{align}
This situation is atypical in that there is a nonlinear encoding procedure from the original qubits to the encoded states. The motivation for exploring such alternative QEC codes lies in getting better performance in terms of quantities such as error thresholds towards fault-tolerant quantum computing. 
  In terms of code-capacity error thresholds (i.e. the error threshold assuming ideal operations), the surface code has been reported to show one of the highest values of up to $ \sim 19 \% $ under depolarizing noise \cite{dennis2002topological,wang2011threshold,bombin2012strong,wootton2012high}. Other codes such as qLDPC  have also shown high thresholds \cite{PhysRevResearch.2.043423,kuo2026degenerate}.    
  Including errors on the operations gives the widely quoted $ \sim 1 \% $ circuit-level threshold for the surface code \cite{fowler2012surface}.   
 }

In this paper, we introduce a novel quantum error correcting code, called the spinor code.  { The spinor code is a nonstabilizer code, as it uses error spaces defined by the total spin.  Furthermore,  it is nonlinear code, where the family of states that are used are spin coherent states (or spinor states) \cite{byrnes2024multipartite}, meaning that it does not rely the traditional encoding in terms of linear superposition of code words.}  { By using the spherical symmetry to treat all single qubit Pauli errors on an equal footing, we propose a QEC procedure that is able to approximately suppress such errors.  We study the effect of various errors on the code and show that Pauli errors are asymptotically correctable in the limit of large code sizes. }  We calculate the code threshold by scaling up the code for a depolarizing channel and find the regimes where the logical error rate can be consistently decreased.

\begin{figure}[t]
\includegraphics[width=\linewidth]{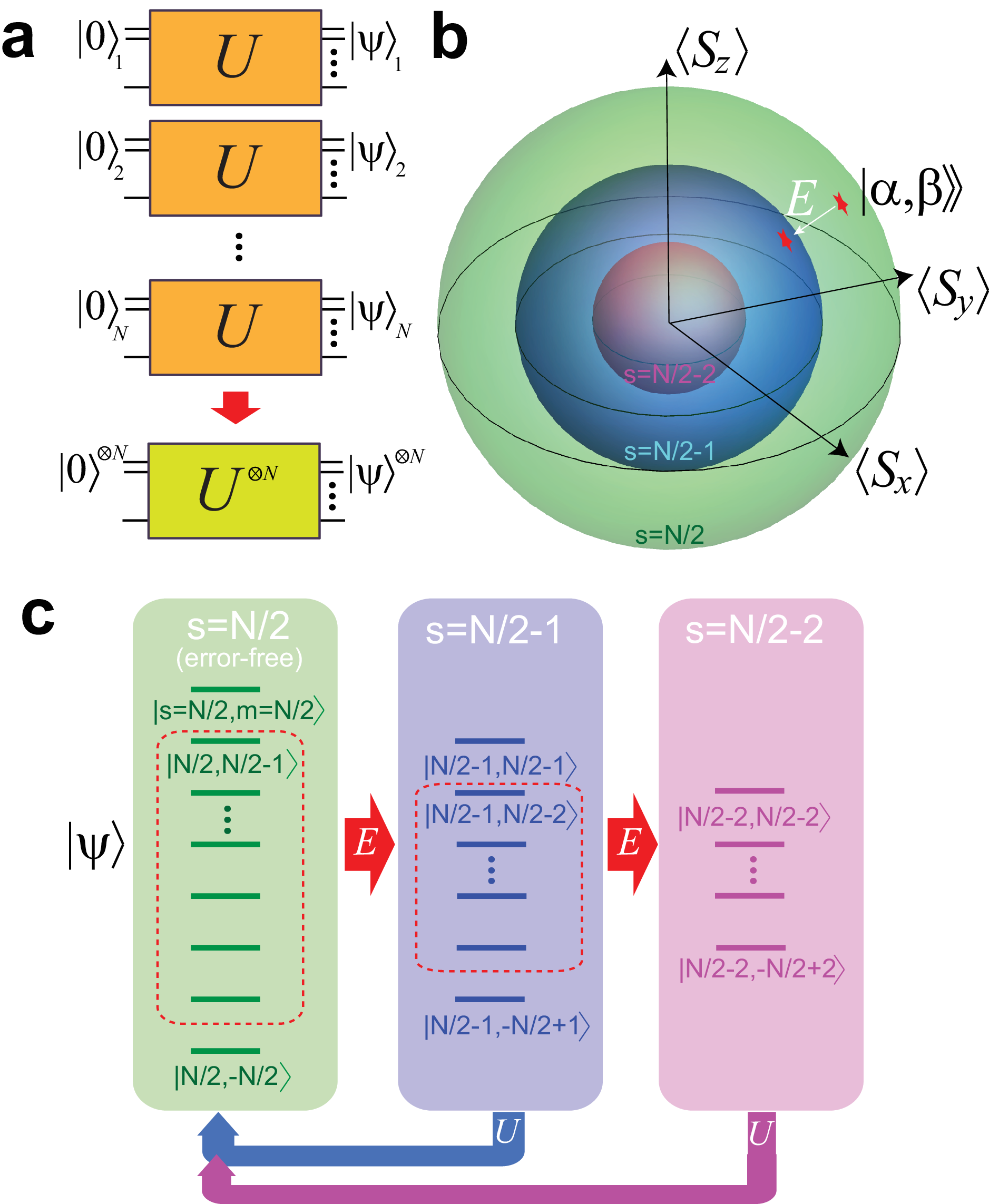}
\caption{(a) $ N $ quantum computers run in parallel, each with $ M $ qubits.  The $m$th qubit starts from the same initial state $ |0 \rangle $ and the quantum circuit  corresponding to the unitary $ U $ is applied, where the final state is unentangled.  After the quantum circuit evolution, the states of the $ m$th qubit of all the quantum computers is $ |\psi \rangle^{\otimes N } $. (b) Code spaces of the spinor code { shown in the Bloch sphere representation}.  The outer sphere (with total spin $ s = N/2 $) is the error-free code space.  Each point on the outer sphere (labeled by the star) corresponds to an encoded qubit state according to (\ref{encoding}).  Single qubit errors acting in the error-free space scatter the state to the inner sphere  with $s = N/2-1 $.  Successive errors events may further scatter the state to inner spheres $ s \rightarrow s \pm 1 $.  {
(c) Code spaces shown in terms of angular momentum eigenstates.  The $ s = N/2 $ sector is the error-free space.  Under ideal errors, the state is mapped to lower spin sectors without deformation. The errors map the states in the dashed boxes due to the dimensional mismatch between spin sectors.  The correction unitary $U_{sl} $ given in (\ref{ucdef}) then maps the state back to the error-free space. } \label{fig1}  }
\end{figure}

\section{The spinor code}

\subsection{Encoding} 

We first define the spinor code and its correction procedure.  We consider an encoding of the quantum information of a qubit as
\begin{align}
{|\psi_0\rangle = } \alpha | 0 \rangle + \beta | 1 \rangle \rightarrow | \alpha, \beta \rangle \rangle { :=}  \left(\alpha | 0 \rangle + \beta | 1 \rangle \right)^{\otimes N }   \label{encoding}
\end{align} 
where $ \alpha, \beta $ are normalized coefficients $ | \alpha |^2 + | \beta |^2 = 1 $, and $ N $ is the number of qubits in the encoded state.  The encoded states in (\ref{encoding}) are spin coherent states, which are equivalent to qubit states duplicated $ N $ times. Some related qubit ensemble encodings have been considered in several works \cite{brion2008error,mohseni2021error,omanakuttan2024fault,omanakuttan2023spin}.  

Such an encoding is often ruled out \cite{knill1997theory} in many discussions of QEC from the outset as it violates the no-cloning theorem \cite{wootters1982single}, which states that there is no physical operation to perform (\ref{encoding}), if the state is unknown.  However, there are situations where such an encoding is physically viable.  To see how this can arise, consider the following example.  Suppose there are $ N $ quantum computers working in parallel, executing the same quantum algorithm, starting from the same initial state (see Fig. \ref{fig1}(a)). Considering the $m$th qubit of the quantum computer, without loss of generality, we may take this to be the $ | 0 \rangle $ state. Taking the $ N $ quantum computers as a single device, we may equally consider it as starting with an ensemble $ | 0 \rangle^{\otimes N } $.  Since the initial state is a known state, the no-cloning theorem is not violated.  Now consider that the the quantum computer proceeds with the calculation and outputs an unentangled state. 
Such a computation may be highly non-trivial, such as with the case of Grover's or Shor's algorithm where the output is a binary state \cite{grover1996fast,shor1994algorithms}.  At the end of the quantum algorithm, the state of the $ m$th qubit across the $N$ quantum computers is $ | \psi \rangle^{\otimes N } $, which is an unknown state due to the complexity of the quantum algorithm.  If there is a procedure to protect the quantum information that is shared by the $ N $ quantum computers such that logical information is better protected against decoherence than the average of the quantum computers individually, then this is a desirable task. We can thus see that in such a scenario, there is no violation of the no-cloning theorem (it is ``pre-cloned''), so that an explicit nonlinear encoder of the form (\ref{encoding}) is not required.

Other more practical applications exist.  A prime application is spinor quantum computing \cite{byrnes2012macroscopic,byrnes2021quantum}, which is a similar approach where the quantum computation proceeds entirely in the completely symmetric subspace \cite{byrnes2024multipartite}.  Another application is quantum metrology with spin ensembles \cite{gross2012spin}, which involve generating spin squeezed states, and uses known states as a resource for precision detection, hence are not affected by the no-cloning theorem.  These examples are further discussed in Appendix \ref{app:spinorqc}.

\subsection{Error spaces}

The Hilbert space of $ N $ qubits can be decomposed in terms of total spin eigenstates $ | s, l, m \rangle $, where the eigenvalues are given by
\begin{align}
S^2 | s, l, m \rangle & = s(s+1) | s, l, m \rangle \label{s2eigs} \\
S_z | s, l, m \rangle & = m | s, l, m \rangle \label{szeigs}
\end{align}
where the total spin operators are $ S_j = \tfrac{1}{2} \sum_{n=1}^N \sigma^{j}_n $, where $ \sigma^{j}_n$ are the Pauli matrices for $ j \in \{x,y,z \} $, and $ S^2 = S_x^2+ S_y^2+ S_z^2$.  Eigenvalues take the values $ s \in \{\tfrac{N}{2}, \tfrac{N}{2}-1, \dots, \tfrac{1}{2}, 0 \} $ and $ m \in [-s, s] $.  $ l \in [1, L_s] $ is a degeneracy label, since there may be more than one spin sector with the same $ s $ eigenvalue (see Appendix \ref{app:totalspin} for further details).  

Spin coherent states have the property that they are maximal spin eigenstates such that $ S^2 | \alpha, \beta \rangle \rangle = \frac{N}{2} (\frac{N}{2} +1 )  | \alpha, \beta \rangle \rangle $.  This can be seen explicitly by writing them in terms of the maximal spin eigenstates
\begin{align}
| \alpha, \beta \rangle \rangle = \sum_{k=0}^N \sqrt{N \choose k} \alpha^k \beta^{N-k} |\frac{N}{2}, 1, k-\frac{N}{2} \rangle .
\label{scsexp}
\end{align}
The maximal spin eigenstates $|\frac{N}{2}, 1, m \rangle$ are completely symmetric states under qubit interchange  { (i.e. Dicke states with $ N/2 - m $ excitations)} \cite{byrnes2021quantum,byrnes2024multipartite}.  

We define the code spaces according to the quantum numbers $ (s,l) $ they possess

\begin{align}
P_{sl} = \sum_{m=-s}^{s} |s,l, m \rangle \langle s,l, m  | ,
\label{psldef}
\end{align}
which are the syndrome measurements that will be performed.   { We note that the span of all possible spin coherent states (\ref{scsexp}) coincides with the span of the states $|\frac{N}{2}, 1, m \rangle$.  Thus }  the error-free code space corresponds to the projection with $ s = N/2, l = 1$.

The basic concept of the spinor code can then be visualized in Fig. \ref{fig1}(b), which shows the Bloch sphere representation for the various spin sectors in an $ N $ qubit encoding.  The error-free encoding of the qubit lies on the outer sphere $ s = N/2 $ of the concentric Bloch spheres.  Errors that occur on the state map the state onto the inner spheres, with total spin $ s < N/2 $. The use of total spin states makes the error spaces spherically symmetric by construction, such that all error types $ j \in \{x,y,z \} $ can be handled on an equal footing.  Then as long as the errors map the states such that the position on the inner sphere is the same to that of the original state, the state is uncorrupted during the error process.  The states are corrected by mapping the states back to the outermost sphere, which completes a single spinor QEC cycle.  In the subsequent sections, we elaborate on the details of this scheme, giving the specific operations for each step.  

{ Our code has similarities with permutation-invariant quantum codes \cite{ruskai1999pauli,pollatsek2004permutationally,ouyang2014permutation,ouyang2026theory}, which also use superpositions of Dicke states as code words. Connections also exist to Ref. \cite{brandao2019quantum}, where codewords are constructed from ground states of many-body Hamiltonians \footnote{In our case, the Hamiltonian would correspond to a Heisenberg ferromagnet, which has ground states $ |N/2, 1, m \rangle $. }.    One major difference to these works is that our encoding (\ref{encoding}) uses a nonlinear manifold of spin-coherent states, rather than a linear code subspace of the symmetric sector. For example, in permutation-invariant codes, a set of logical states are chosen in the symmetric sector, and the encoding is a conventional linear superposition of them, e.g. $ \alpha |0_L \rangle + \beta |1_L \rangle $. Here, we use the full set of states $|\frac{N}{2}, 1, m \rangle$, but only consider states taking the specific form (\ref{scsexp}).  }



\subsection{Decoding}

In order to extract information from the encoded state, in QEC one must perform a decoding operation to extract the stored information. { While our encoding (\ref{encoding}) is nonlinear, the decoding process can be written as a (linear) completely positive trace preserving (CPTP) map
\begin{align}
{\cal D} (\rho) = \frac{1}{2} \left[ I +   \frac{2}{N} \sum_{j=1}^3  \langle S_j \rangle \sigma^j  \right] .
\label{decodedstate}
\end{align}
This takes advantage of the correspondence between spin coherent states and the equivalent qubit state
}
\begin{align}
    \frac{\langle S_j \rangle}{N/2} \leftrightarrow \langle \sigma^j \rangle 
    \label{qubitscsmap}
\end{align}
for $ j \in \{x,y,z \}$. For a spin coherent state, this reproduces exactly the same expectation values as for the original qubit \cite{byrnes2021quantum}.

{
\subsection{Logical error}
}

The logical error of the state { after the} decoding  { is evaluated using the trace distance
 \begin{align}
    \begin{split}
    \epsilon_\text{L} & =  \frac{1}{2} \Big| \Big| {\cal D} (\rho) - | \psi_0 \rangle \langle \psi_0 |  \Big| \Big| \\    
    & = \frac{1}{2} \sqrt{ \sum_{j=1}^3 ( \text{Tr} [ {\cal D} (\rho) \sigma_j ]  - \langle \psi_0 | \sigma_j | \psi_0 \rangle  )^2 } 
    \label{logicalerror}
     \end{split}
\end{align} 
where the initial state $ | \psi_0 \rangle $ before encoding is given in 
(\ref{encoding}).  We note that for our nonlinear code, it is important to perform the decoding before evaluating the logical error (see Appendix \ref{app:logicalerror}).  
}

\section{Idealized errors}
\label{sec:ideal}

{
In this section, we show that the code spaces defined in (\ref{psldef}) form a valid QEC code.  To simplify the argument, we first consider a set of idealized errors, which cleanly (i.e. without deformation) map the states between the error-free space to the error spaces.  This will allow us to highlight several features of the code spaces (\ref{psldef}), as well as defining a correction operation.  We also directly show that the Knill-Laflamme criterion is satisfied explicitly, meaning that for the idealized errors, the code spaces (\ref{psldef}) can form a conventional linear QEC code.  This highlights the fact that the nonlinear encoding (\ref{encoding}) is only strictly necessary when considering more realistic Pauli errors as considered later.  
}

First consider errors of the type
\begin{align}
F_{s l \tilde{l} } & = \sqrt{p_{sl \tilde{l} }} \exp \left[ i \frac{\pi}{2} \sum_{m=-s}^{s} \left(  | s, l, m \rangle \langle s+1, \tilde{l}, m | + \text{H.c.}  \right)  \right]  .
\label{protoerror}
\end{align}
This corresponds to an error that scatters the state from the $ (s + 1, \tilde{l}) $ sector to $ (s,l) $ and vice versa { (see Fig. \ref{fig1}(c)). }  The error preserves the $ m $ eigenvalue of the state. We consider (\ref{protoerror}) to be an ``ideal'' error with respect to the spinor code, which can be perfectly corrected as long as there is only one error. We will see later that single qubit $ \sigma^z $ errors have similarities to this type of error. This satisfies
\begin{align}
\sum_s \sum_{l=1}^{L_s}  \sum_{\tilde{l} =1}^{L_{s+1}}  F_{s l \tilde{l}}^\dagger F_{s l \tilde{l}} = I ,
\end{align}
for a probability distribution $ \sum_s \sum_{l=1}^{L_s}  \sum_{l=1}^{L_{s+1}} p_{sl \tilde{l} } = 1 $. 

Evaluating the Knill-Laflamme QEC criterion \cite{knill1997theory,nielsen2002quantum}, we have
\begin{align}
    \langle C_m | F_{q}^\dagger F_{q'} | C_{m'} \rangle = h_{q q'} \delta_{mm'} ,
    \label{klcrit}
\end{align}
where we defined $ q =(s,l, \tilde{l}) $ and $ q' =(s',l', \tilde{l}' ) $ and the { error-free code space basis state} as $ |C_m \rangle = | \frac{N}{2}, 1, m \rangle $ for the range $ m \in [ -m_{\max}, m_{\max} ] $. We also defined the matrix elements $   h_{q q'} =  \sqrt{p_{q}p_{q'}} g_{q q'} $,  where
\begin{align}
g_{q q'} = \left\{ 
\begin{array}{cc}
1 & \text{ if } s, s' < N/2-1 \\
0 &  \text{ if } s= N/2 -1 \text{ and }  s' < N/2-1  \\
0 &  \text{ if } s < N/2 -1 \text{ and }  s' = N/2-1  \\
\delta_{ll'}  &  \text{ if } s,s' = N/2 -1 
\end{array} 
\right. .
\end{align}
We see that (\ref{klcrit}) satisfies the Knill-Laflamme conditions for the QEC code, since $ h_{q q'}  $ is a Hermitian matrix independent of $ m $.

The correction procedure works in the following way.  The error (\ref{protoerror}) has the effect of mapping the states in sector $ (s+1, \tilde{l}) $ to the sector $ (s, l) $  { (see Fig. \ref{fig1}(c)).}  As the initial state starts in the maximally polarized sector, one application of the error maps the state from the $ s = \frac{N}{2} \rightarrow \frac{N}{2} -1 $ sector. 
By performing a syndrome measurement (\ref{psldef}) we may determine which spin sector $ (s, l )$ the state has ended up in.  The correction operation should be taken as 

\begin{align}
U_{sl} = \exp \left[  i \frac{\pi}{2} \sum_{m=-s}^s \left(  |\frac{N}{2}, 1, m \rangle \langle s, l, m| + \text{H.c.} \right) \right] .
\label{ucdef}
\end{align}
This unitarily rotates the states from the $ (s, l) $ subspace to the { error-free space} $ s = N/2 $, preserving the $ m $ eigenvalues.  

The above procedure will correct any state of the form 
\begin{align}
| \psi \rangle = \sum_{m=-m_{\max}}^{m_{\max}}  \psi_m | C_m \rangle
\label{idealstatekl}
\end{align}
from the errors (\ref{protoerror}), where $ \psi_m $ are normalized coefficients. In this case, we have a standard linear QEC. We will see in the next section that for more realistic errors the mapping between spin sectors is not as clean, and some deformation of the state takes place.  However, by restricting the types of states that we intend to correct to spin coherent states (\ref{scsexp}), a similar procedure can be performed to counteract errors.  This converts the linear QEC to a nonlinear QEC, as a restricted family of states is considered correctable.  

The reason for the restriction in range $ m \in [ -m_{\max}, m_{\max} ] $ is that for successive applications of $ F_{q} $, the state has the possibility of being scattered to smaller spin sectors $ s = \frac{N}{2} \rightarrow \frac{N}{2} -1 \rightarrow \frac{N}{2} -2 \dots $ { (see Fig. \ref{fig1}(c)).}  Since the maximum range of smaller spin sectors is restricted $ m \in [-s,s] $, the projection (\ref{psldef}) can potentially cause truncation of the state near the poles of the Bloch sphere.  Thus by working within a range $ m \in [ -m_{\max}, m_{\max} ] $, this truncation is avoided up to a certain number of error events.  The cutoff defines the maximum spin where there is the full range $ s = m_{\max} $.  This therefore defines the code distance, defined as the number of error events the code can hangle before a logical error occurs.  The code distance for the spinor code is
\begin{align}
d = \frac{N}{2} - m_{\max}  .
\label{codedistance}
\end{align}
We see that increasing the number of qubits $ N $ generally increases the code distance.  

Going forward, we will consider more realistic Pauli errors, which have a similar action as the idealized errors (\ref{protoerror}), where the total spin can only change by one unit $ s \rightarrow s \pm 1 $.  We will also 
relax the limitation where the code space should be limited in range $ m \in [ -m_{\max}, m_{\max} ] $, and simply consider the full range $ m \in [-N/2, N/2] $. 
In this situation, we expect states near the poles to be more susceptible to errors but those near the equator to be able to withstand more errors. 
We expect for a large number of qubits $ N $, the effect of these missing states to become less important, as the dimension mismatch becomes a smaller fraction of the code dimension.  { Specifically, the error-free space has a dimension of $ N + 1 $, compared to the single Pauli error space of $ s = N/2-1 $, which has a dimension of $ N-1 $.  }
This is in contrast to codes such as the bit flip code, where beyond the code distance number of errors, the state suffers a dramatic logical error.  Here, the introduction of logical errors is more gradual.

\section{Pauli errors}

We now consider some more realistic errors, and analyze its effect on the code.  In this section, we will consider single qubit Pauli errors from two viewpoints.  The first will be to directly evaluate the effect of such errors on the code space. The second will be to show that the Knill-Laflamme conditions are approximately satisfied under particular assumptions. { We note this section is concerned with the effect of single errors; multiple errors will be considered in the next section.}

\subsection{Deformation factors}

\subsubsection{Phase flip errors}
\label{sec:phaseflipdeform}

To understand the way in which errors act on the spinor code, let us evaluate the effect of single qubit errors on a general state in the error-free code space after the syndrome measurement.   Considering $ \sigma^z $-errors first, we have  
\begin{align}
P_{sl} \sigma^z_n | \psi \rangle = \sum_{m=-s}^s \psi_m D_{sl}^{(n)} (m) | s, l, m \rangle ,
\label{deformz}
\end{align}
where $ \psi_m = \langle \tfrac{N}{2}, 1, m | \psi \rangle $ are the original amplitudes. The factors 
\begin{align}
D_{sl}^{(n)} (m)  = \langle s, l, m | \sigma_n^z  |  s=\tfrac{N}{2}, 1, m \rangle \label{deformationdef}
\end{align}
are {\it deformation factors} since they show the effect on the state following a syndrome measurement when an $ \sigma^z $-error has occurred. For the state to be mapped without deformation to the error space, $ D_{sl}^{(n)} (m) $ should be a constant in $ m $.  We note that $ \sigma^z $ errors cannot change the $ S^z $ eigenvalue, which is why only diagonal matrix elements in $ m $ appear in (\ref{deformationdef}).  

By direct evaluation, we find deformation factors follow 
\begin{align}
D_{sl}^{(n)} (m) = \left\{
\begin{array}{ll}
\frac{m}{N} & s=\tfrac{N}{2} \\
A_{l}^{(n)} (1+ b_s \frac{m^2}{N^2} + c_s \frac{m^4}{N^4} ) & s=\tfrac{N}{2}-1 \\
0 & \text{otherwise}
\end{array}
\right. . \label{dfactors}
\end{align}
Here, $ A_{l}^{(n)} $ is an $ m$-independent function that satisfies $ \sum_{l=1}^{L_s} | A_{l}^{(n)}  |^2 = 1$.  We also have $ \sum_{s=0}^{N/2} \sum_{l=1}^{L_s} | D_{sl}^{(n)} (m) |^2 = 1$.  We emphasize that the factors (\ref{dfactors}) are exact in the sense that the functional dependence on $ m $ is completely described up to the coefficients $ A_{l}^{(n)}, b_s, c_s $.  

\begin{figure}[t]
\includegraphics[width=\linewidth]{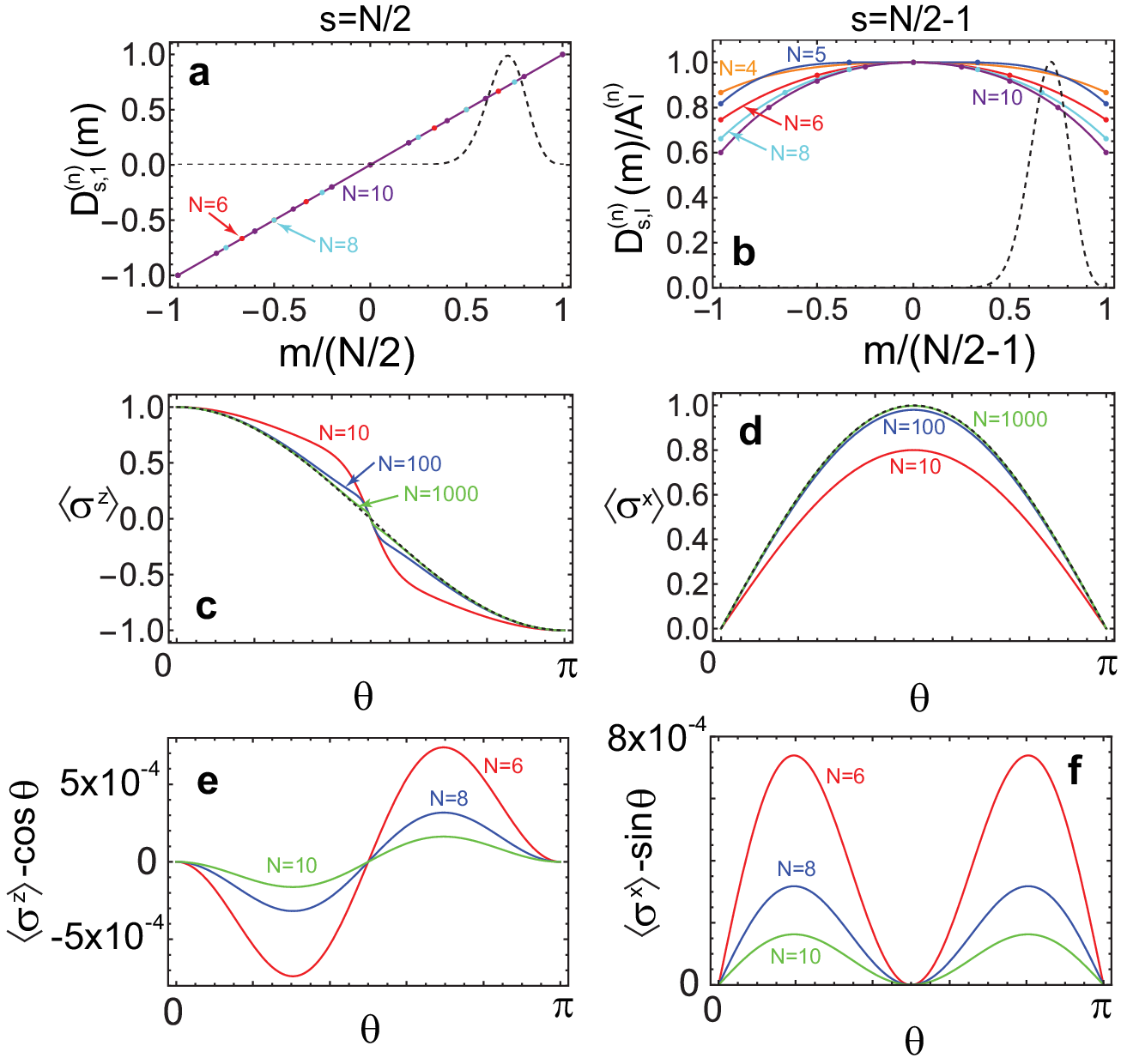}
\caption{{ Effect of Pauli errors on the spinor code.}
(a)(b) Deformation factors (\ref{deformationdef}) for various $ N $. Spin sectors are (a) $ s = N/2 $ and (b) $ s = N/2 - 1 $.  Dots correspond to directly evaluated values according to the definition (\ref{deformationdef}).  Solid lines are (a) $ m/N $ and (b) $ 1 + b_s (m/N)^2 + c_s (m/N)^4 $ as given in (\ref{dfactors}).  The dashed line is the spin coherent state amplitude $ \psi_m = \langle \tfrac{N}{2}, 1, m | \alpha, \beta \rangle \rangle $ for $ N = 100$ and $ \alpha = \cos \tfrac{\pi}{8}, \beta = \sin \tfrac{\pi}{8} $. All plots are independent of both the qubit number $ n $ and degeneracy number $ l $. { (c)(d) Decoded spin expectation values for the state (\ref{deformz}) with $ s = N/2, l = 1 $ starting the spin coherent state (\ref{scsexp}).  This is then decoded using (\ref{decodedstate}) and the expectation values (c) $ \langle \sigma^z \rangle $ for all $ \phi $ and (d)  $ \langle \sigma^x \rangle $ for $ \phi = 0 $.  Dashed lines correspond to expectation values of the state with no error. (e)(f) The same as (c)(d) but with $ s = N/2 -1 $ and any $ l $.  Due to the small difference with the zero error curves, we plot the difference (e) $ \langle \sigma^z \rangle - \cos \theta $; (f) $ \langle \sigma^x \rangle - \sin \theta \cos \phi$ for $ \phi = 0 $.  }
\label{fig2}  }
\end{figure}

Clearly, the deformation factors (\ref{dfactors}) are not independent of $ m $ hence we cannot expect that the spinor code to be an exact QEC under Pauli errors.  However, the $ m $ dependence is not very strong and for particular types of states its effects can be small to negligible (see Fig. \ref{fig2}). In Fig. \ref{fig2}(a) we show the deformation factor for projections into the same $ s = N/2 $ sector. Since a spin coherent state has a Gaussian distribution in $ m $, the effect of the deformation factor is to displace the distribution towards the extrema $ m = \pm N/2 $.  
{ 
This can be quantified by evaluating the expectation values of the state (\ref{deformz}) with respect to the spin coherent state (\ref{scsexp}).  For a spin coherent state with parameters $ \alpha = e^{-\i \phi/2} \cos (\theta /2), \beta = e^{\i \phi/2} \sin (\theta /2) $, we find the decoded state (see Appendix \ref{app:deformation})
\begin{align}
\langle \sigma^x \rangle & = \sin \theta \cos \phi - \frac{2}{N} \sin \theta \cos \phi \nonumber \\
\langle \sigma^y \rangle & = \sin \theta \sin \phi - \frac{2}{N} \sin \theta \sin \phi \nonumber \\
\langle \sigma^z \rangle & = \cos \theta + \frac{2}{N} \frac{\sin^2 \theta}{\cos \theta} + O(\frac{1}{N^2} ) 
\label{logicalshift}
\end{align}
where the expectation values are taken with respect to the decoded state (\ref{decodedstate}) using the projection to $ s= N/2, l = 1 $ in (\ref{deformz}).  We see that the errors in the original positions diminish as $ O(1/N) $, due to the fact that the spin coherent state 
}
Gaussian has a width of approximately $ \Delta m/(N/2) \sim 1/\sqrt{N} $.  Hence we expect that for large $ N $ the spinor code to improve in performance, { on a per error basis. Figure \ref{fig2}(c)(d) shows a plot of expectation values $ \langle \sigma^j \rangle $ for the decoded state.  We see that as expected, for large $ N $ the states are largely unaffected by single phase flip errors.  The strongest deformations are in the vicinity of $ \theta = \pi/ 2$.  However, the probability of these events are suppressed since the deformation factor also has a small amplitude near $ m = 0 $.}  For spin coherent states at the extrema $ m = \pm N/2$, the Gaussian narrows and $ D^{(n)}_{s,1} (m) \approx \pm 1 $.  In this region the errors do not affect the state at all.  This occurs as $z$-errors do not affect the states at the poles of the Bloch spheres, as they are $S_z$-eigenstates.

{ 
The above considers the effect of a single error on the spinor code.  As the projection step does not perfectly revert the state back to a spin coherent state, it is important to understand how the state evolves for subsequent errors.  To analyze this, it is illustrative to consider the worst case in terms of the deformation of the state, corresponding $ \theta = \pi/2 $, where the resultant state (\ref{deformz}) is changed the most.  Taking the case of $ \phi = 0 $, the initial state prior to the error is 
\begin{align}
    | \frac{1}{\sqrt{2}}, \frac{1}{\sqrt{2}} \rangle \rangle = | + \rangle^{\otimes N} = | \frac{N}{2}, 1, \frac{N}{2} \rangle^{(x)} ,
\end{align}
where we defined }
\begin{align}
    |s, l, m \rangle^{(x)}  = e^{-i S^y \pi/2 } |s, l, m \rangle 
\end{align}
and the total spin eigenstates in the $ x $-basis.  These satisfy 
\begin{align}
S^2 | s, l, m \rangle^{(x)} & = s(s+1) | s, l, m \rangle^{(x)} \\
S_x | s, l, m \rangle^{(x)} & = m | s, l, m \rangle^{(x)} .
\end{align}
{
After a phase flip error $ \sigma^z$ and projection we have
\begin{align}
P_{sl} \sigma^z_n | \frac{N}{2}, 1, \frac{N}{2} \rangle^{(x)} = 
\frac{1}{\sqrt{N}} | \frac{N}{2}, 1, \frac{N}{2}-1 \rangle^{(x)} .
\end{align}
We see that the phase flip error shifts the $ m $ by one unit in the $ x $-basis. Repeating this, it is clear that subsequent errors then shift the errors by $ \Delta m = \pm 1 $.  Hence we expect that under multiple errors, the states will undergo diffusion away from their initial values.  Since a step of $ \Delta m = \pm 1 $ corresponds to a shift of $ \sim 1/N $ in terms of expectation values, we expect that the general behavior of an $ O(1/N) $ shift per error to hold under multiple errors.  In Sec. \ref{sec:threshold} we will explicitly study the effect of multiple errors. }

For projections to the $ s = N/2 - 1 $ sector, the effect on spin coherent states are even weaker.  First, the amplitude factors $ A_{l}^{(n)} $ appearing in (\ref{dfactors}) are irrelevant in terms of how the state is deformed, since they are $ m $-independent and do not affect the state (\ref{deformz}).  The magnitudes of these coefficients affect the probability of obtaining the outcome $ (s, l ) $ in the syndrome measurement.  Around $ m = 0 $ the distribution of the deformation factors is typically rather flat and symmetric.  As such, for spin coherent states in the vicinity of $ m = 0 $ (near the equator of the Bloch sphere), the distributions are virtually unaffected.  Thus in these regions the states are mapped to the error space with little deformation.  Near the extrema, the the deformation factors act to shift the Gaussian towards $ m = 0 $.  However, again, as $ N $ is increased the effect diminishes since the Gaussian tightens as $ \Delta m/(N/2) \sim 1/\sqrt{N} $. { The spin expectation values $ \langle \sigma^i \rangle $ of the decoded state are plotted in Fig. \ref{fig2}(e)(f).  We see that even for the relatively small $ N $ values that are calculated, the deviations from the zero error state are very small, and furthermore diminishing with $ N $.  This confirms that for projections into the $ s = N/2 - 1$ sector, the deformation factors have a minimal effect. 
}


We note that single qubit errors do not create a transition from $ s = N/2 $ to lower spins $ s < N/2 - 1$, as can be seen from (\ref{dfactors}).   Hence the $ s = N/2 $ and the $ s = N/2 - 1$ are the only spaces that need to be considered.  { There is a similar effect exploited in permutation-invariant codes where single Pauli errors create transitions out of the completely symmetric sector \cite{ouyang2014permutation,ouyang2026theory}. }With multiple errors, of course it is possible for the states in the  $ s = N/2 - 1$ sector to transition down to  $ s = N/2 - 2$ and so on.  For this reason, we expect the spinor code to be an approximate nonlinear QEC for Pauli errors, improving as $ N $ is increased.   We note here that the way that $ z$-errors occurs is similar to the idealized errors (\ref{protoerror}).  In both cases the $ m $ quantum number is unchanged and the spin $ s $ only changes by one unit.  The primary difference, in the context of the spinor code, is the presence of the deformation factors which affect the distribution of the states.

{ We thus conclude that on a per error basis, }
the effect on the errors on spin coherent states are modest and diminish with $ N $. {
We however note that as $ N $ is increased, the prevelance of single qubit errors also increases as $ N $. Thus although increasing $ N $ generally has an $ O(1/N) $ effect on the encoded state, this diminishing effect is countered by a factor $ N $ increase in probability of the error occurring. This requires an analysis that takes into account of multiple errors, which we study further in Sec. \ref{sec:threshold}. 
}

\subsubsection{Bit flip errors}
\label{sec:bitflip}

Now let us consider errors other than phase-flip errors.  Considering bit flip errors, we have
\begin{align}
P_{sl} \sigma^x_n | \psi \rangle = \sum_{m=-s}^s \psi_m^{(x)} D_{sl}^{(n)} (m) |s, l, m \rangle^{(x)}  
\label{xdeform}
\end{align}
where $  \psi_m^{(x)} = \langle \tfrac{N}{2}, 1, m | \psi \rangle $ are the amplitudes in the $ x $-basis. In (\ref{xdeform}), we used the fact that the spin projections can be equally written in the $ x $-basis
\begin{align}
P_{sl} = \sum_{m=-s}^{s} |s,l, m \rangle^{(x)}  \langle s,l, m  |^{(x)}  .
\end{align}
since the $ x $-basis states completely span the $ (s, l) $ subspace.  

We observe in (\ref{xdeform}) that bit flip errors have exactly the same relation as phase flip errors, except that the deformation factors add in the $ x $-basis.  Note that the deformation factors appearing in (\ref{xdeform}) are identical to the phase flip case.  This illustrates the advantage of using total spin subspaces to encode the quantum information, which possess a spherical symmetry in terms of the subspaces that are used.  In this way, qubit errors in any direction can be treated on the same footing.

\subsubsection{$Q$-functions}

\begin{figure}[t]
\includegraphics[width=\linewidth]{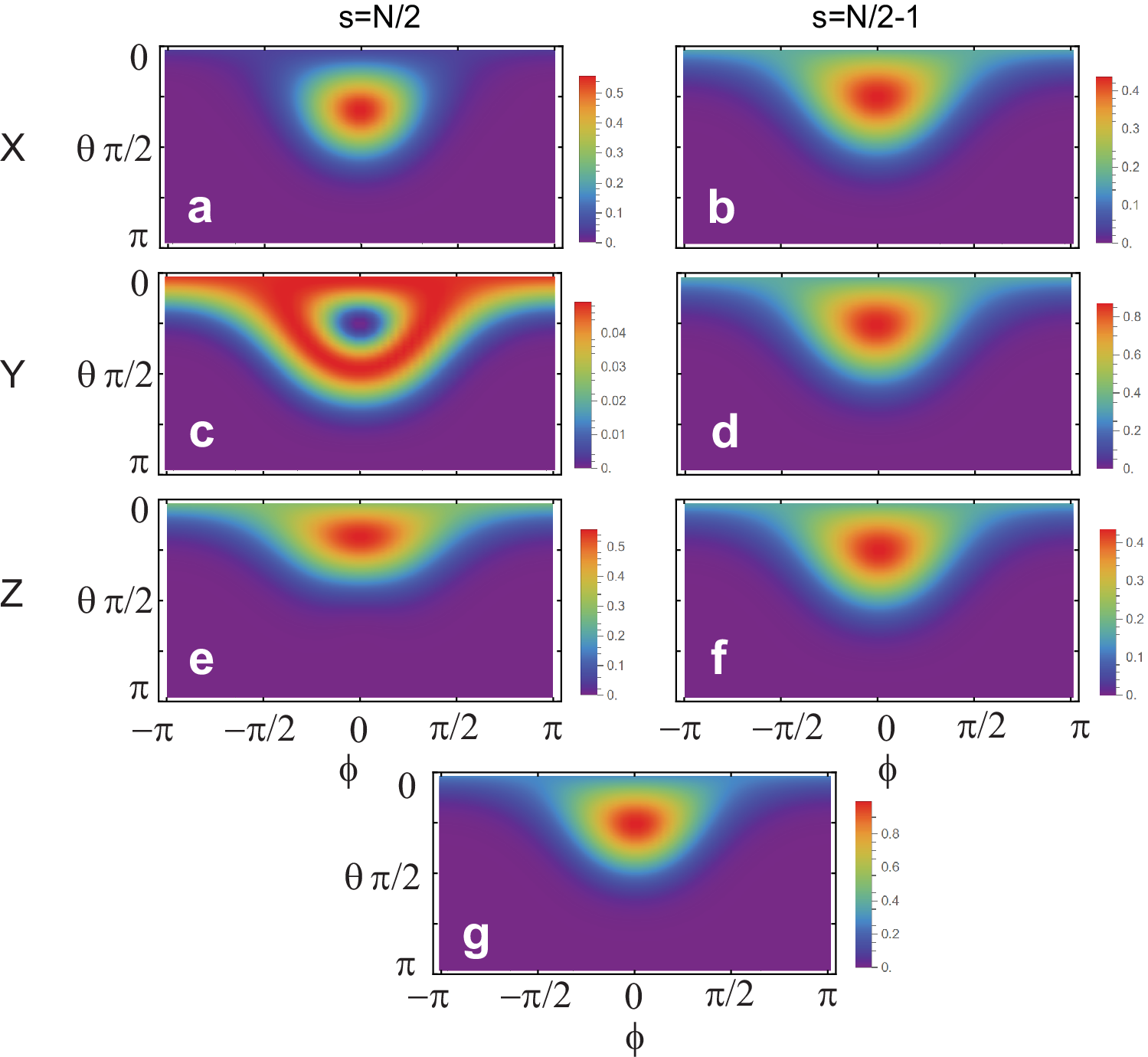}
\caption{$Q$-functions of the projected state (\ref{errorscs}).  The $ Q $-functions are defined as $ Q(\theta, \phi) = | \langle \psi^j_{sln} | \cos \tfrac{\theta}{2}, e^{i \phi} \sin \tfrac{\theta}{2} \rangle \rangle |^2 $.  The left column corresponds to projections in the $ s= N/2 $ sector, while the right column corresponds to projection to the $ s = N/2 -1, l = N $ sector. The rows show errors in the $ j =x,y,z $ directions respectively.  (a) $ j = x, s=N/2, l=1 $; (b) $ j = x, s=N/2-1, l=N $; (c) $ j = y, s=N/2, l=1 $; (d) $ j =y, s=N/2-1, l=N $; (e) $ j = z, s=N/2, l=1 $; (f) $ j = z, s=N/2-1, l=N $.  (g) $Q$-function of the original state $ Q(\theta, \phi) = | \langle \langle \alpha, \beta | \cos \tfrac{\theta}{2}, e^{i \phi} \sin \tfrac{\theta}{2} \rangle \rangle |^2 $.  Common parameters are $ \alpha = \cos \tfrac{\pi}{8}, \beta = \sin \tfrac{\pi}{8}, N = 8 , n = 1 $.  
\label{fig3}  }
\end{figure}

Finally, to illustrate the effect of various errors acting on spin coherent states, we plot the $Q$-functions of the state
\begin{align}
    |\psi^j_{sln} \rangle = P_{sl} \sigma^j_n | \alpha, \beta \rangle \rangle ,
    \label{errorscs}
\end{align}
which is a spin coherent state with an error in the $ j$-direction, projected to the $ (s, l) $ subspace.  The left column of Fig. \ref{fig3} shows the $ Q $-functions for errors in the $ j = x,y, z $ directions respectively projected to the $ s = N/2 $ subspace.  For the $ \sigma^x $ and $ \sigma^z $ errors (Fig. \ref{fig3}(a)(e)), we see the effect predicted by the distribution in Fig. \ref{fig2}(a):  the distribution moves slightly towards the $x$ and $ z $ axes due to the modification of the Gaussian distribution by the linear deformation factor (Fig. \ref{fig3}(g) is the original distribution).   For the $  y $-error we see a more dramatic change in the distribution, where a Fock state-like distribution is seen \cite{gerry2023introductory}.  This occurs due to the fact that the $ y $-axis is oriented at an angle $ \pi/2 $ with respect to the state, and hence in the $ y $ basis, the Gaussian is centered at $ m_y = 0 $.  The amplitude of the $ Q $ function is much smaller, implying a smaller probability, as predicted previously.  Furthermore, the center of the distribution is unaffected, which suggests that spin averages will not be displaced.  For the projections to the $ s = N/2 - 1$ subspace, there is almost no change to the distribution, as predicted by the deformation factors in Fig. \ref{fig2}(b).  This demonstrates explicitly that the spinor code approximates a quantum error correcting code for Pauli errors.

\subsection{Knill-Laflamme conditions}
\label{sec:knilllaflammeconds}

{
We now analyze the effect of Pauli errors from a different point of view, using the Knill-Laflamme conditions.  One issue with the Knill-Laflamme conditions is that it assumes a linear QEC code, whereas we have a nonlinear encoding (\ref{encoding}). Here, we adapt the Knill-Laflamme conditions to additionally include the additional assumption that only states of the form (\ref{encoding}) need to be protected.  In this section we first present a simple version of the argument, followed by a generalized version in Appendix \ref{app:new_akl}.

}
We consider errors of the form
\begin{align}
E^{(n)}_j = \left\{
\begin{array}{cc}
\sqrt{1-p} I & \text{if } j = 0 \\
\sqrt{p} \sigma^z_n & \text{if } j = 1 
\end{array}
\right. ,
\label{klerrors}
\end{align}
where $ p $ is the single qubit error probability.  Evaluating the Knill-Laflamme QEC criteria with respect to the { error-free code space basis state} $ | C_m \rangle = | \tfrac{N}{2}, 1, m \rangle  $ as before, we obtain 
\begin{align}
\langle C_m | E_j^\dagger E_{j'} | C_{m'} \rangle = w_{j j'} \delta_{m m'}
\label{knilllaflammepauli}
\end{align}
where the matrix $ w$ is
\begin{align}
w = 
\left(
\begin{array}{cc}
1- p  & a_m \\
a_m & p  \\
\end{array}
\right) ,
\label{wmatrix}
\end{align}
where
\begin{align}
a_m &  = \sqrt{p(1-p)} \frac{m}{N} .
\label{amdef}
\end{align}

From the form of (\ref{knilllaflammepauli}) we can see that in general the Knill-Laflamme conditions are not strictly satisfied due to the $ m $-dependence of the matrix $ w $. The $ m $-independence is important because diagonalizing the $ w $ matrix defines new operators
\begin{align}
F_j = \sum_{j'} u_{j j'} E_{j'}  
\label{orthoerrors}
\end{align}
which define the orthogonal error spaces corresponding to different types of errors \cite{nielsen2002quantum}. Here $  u  $ is a unitary matrix which diagonalizes $ w $.  If the $ F_j $ errors are $ m $-dependent, then a consistent error correction procedure cannot be defined for a general superposition state of { $  | C_{m} \rangle $.  }

However, under particular assumptions, we can show that the Knill-Laflamme conditions are approximately satisfied. { In the spinor code we only consider encoding states of the form (\ref{scsexp}).  Spin coherent states have a dominant amplitude in a band 
\begin{align}
   m \approx N( |\alpha|^2 - |\beta|^2) \pm \sqrt{N} .
   \label{mspread}
\end{align}
Hence for large $ N $ we may approximate $ m /N $ by a constant. In Appendix \ref{app:klcond}, we show that for errors satisfying
\begin{align}
   1 & \ll \frac{|1-2p|}{\sqrt{p(1-p)}}    \label{klcriterion} 
\end{align} 
} all the eigenvectors and eigenvalues of (\ref{wmatrix}) become $ m $-independent to a good approximation.  In this regime the Knill-Laflamme conditions can be satisfied, such that a valid QEC code can be formed.  

{
Let us understand what this means in practice. Under condition (\ref{klcriterion}), for a spin coherent state the $ w $-matrix can be taken to be approximately constant, due to the limited range in $ m $, as given in (\ref{mspread}).  Then on application of the $ \sigma^z_n $ error on the state (\ref{scsexp}), the effect on the state is approximately to multiply by the factor $ m/N $, which is taken to be a constant.  This does not affect the state to a good approximation for $ N \gg 1 $. In this case, there is only one code space, and the correction operation is the identity operation. We see that this  approach completely omits the existence of other code spaces with $ s < N/2 $, but it is consistent with the projection outcome $ s = N/2 $, as discussed in Sec. \ref{sec:phaseflipdeform}.  We add that the same deformation factor appearing in (\ref{deformationdef}) appears in (\ref{amdef}).  }



For other types of errors, such as bit flips errors,  the same arguments can be repeated using the { error-free code space basis states} defined along other orientations, similarly to Sec. \ref{sec:bitflip}. In Appendix \ref{app:new_akl}, we also show more formally that the Knill-Laflamme conditions are satisfied approximately for a depolarizing channel.  In this way, we expect that the spinor code approximately can protect against Pauli errors along any axis.

\section{Error threshold}
\label{sec:threshold}

We now directly evaluate the performance of the spinor code.  We will consider specifically the depolarizing channel acting on the $n$th qubit with Kraus operators
\begin{align}
E^{(n)}_j = \left\{
\begin{array}{cc}
\sqrt{1-p} I & \text{if } j = 0 \\
\sqrt{\frac{p}{3}} \sigma_n^j & \text{if } 1\le j \le 3 
\end{array}
\right. 
\label{krausdepol}
\end{align}
where $ p $ is the physical error probability, and $I $ is the identity matrix.  In a single round, this error channel is applied to all $ N $ qubits in the ensemble. 
This way of applying the errors ensures that there is a hierarchy of errors such that with probability $ (1-p)^N $ there are no errors in the system, with probability $ \sim p(1-p)^{N-1} $ there is one Pauli error, with probability $ \sim p^2(1-p)^{N-2} $ there are two Pauli errors, and so on. 
After the error channel is applied, the state is then corrected using the spinor code QEC procedure.  

Concretely, our procedure follows the following sequence:
\begin{enumerate}
    \item[0)] Initialize the state in $ \rho = | \alpha, \beta \rangle \rangle \langle \langle \alpha, \beta | $. 
   \item[1)] Apply the depolarizing error channel 
   \begin{align} \rho \rightarrow \sum_{j=0}^3 E_j^{(n)} \rho (E_j^{(n)})^\dagger 
   \end{align}
   for all $ n \in [1,N] $. 
 \item[2)] Perform the error syndrome measurement and correction
\begin{align}
\rho \rightarrow \sum_{s=0}^{N/2} \sum_{l=1}^{L_s}  U_{sl} P_{sl} \rho P_{sl}^\dagger U_{sl}^\dagger . 
\label{projcorrect}
\end{align}
 \item[3)] { Decode the state using (\ref{decodedstate}) and } measure the logical error (\ref{logicalerror}).
  \item[4)] Go to step 1.  
\end{enumerate}

\begin{figure}[t]
\includegraphics[width=\linewidth]{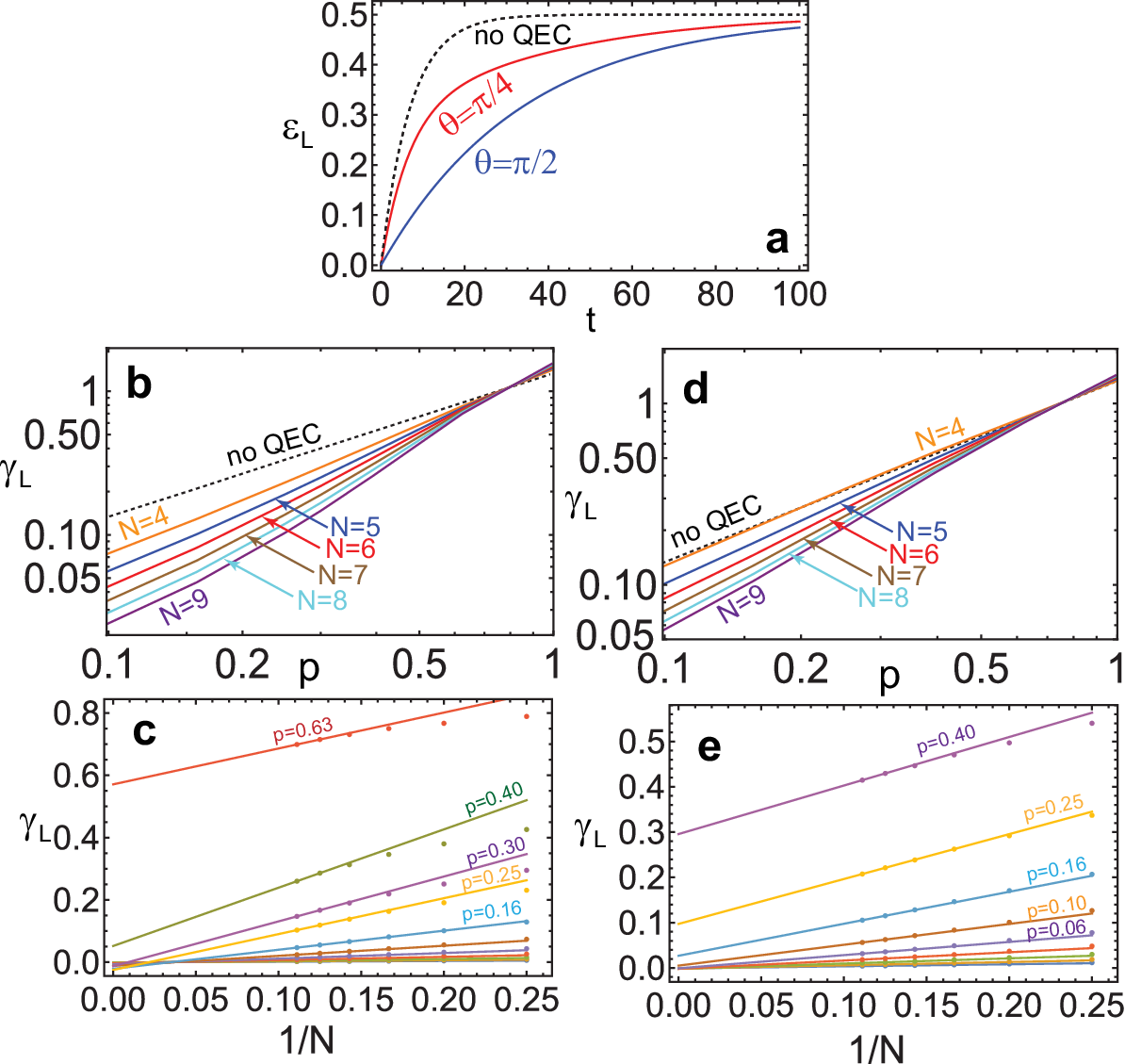}
\caption{Performance of the spinor code.  (a) Logical error  (\ref{logicalerror}) as a function of the number of cycles with a physical error rate $ p = 0.1$ and $ N = 8 $ qubits for various initial states  $ \alpha = \cos \tfrac{\theta}{2}, \beta = e^{i \phi} \sin \tfrac{\theta}{2}$ as marked (solid lines).  There is no $ \phi $ dependence to all curves.  Dotted line shows the logical error skipping the error detection and correction step (step 2) in the sequence.  The same curve is obtained for all initial states for no QEC.  (b) Logical error rate $ \gamma_\text{L} $ as a function of physical error probability for various code sizes $ N $. The logical error rate is estimated by finding the derivative $ \gamma_\text{L} \approx 2 (\epsilon_\text{L} (t=1) -  \epsilon_\text{L} (t=0))$.  The initial state is chosen as $ \theta = \pi/2, \phi = 0 $.  Dotted lines shows the result skipping error detection and correction for all $ N $.  (c) Same data as (b) but plotted with $ 1/ N $ for constant physical error probability.  Points are calculated data, lines are linear fits to the largest two $ N $ values.  (d)(e) Same as (b)(c) but including initialization and measurement errors $ p_{\text{m}}= p_{\text{i}} = p $. 
\label{fig4}  }
\end{figure}

The result of applying many such cycles of errors and QEC is shown in Fig. \ref{fig4}(a).  Here we plot the logical error as a function of the number of cycles $ t $ of our sequence for various initial states with and without performing the QEC.  For the case with no QEC, the procedure is the same as above but we omit step 2 in the above sequence.  All initial states for the ``no QEC'' case give the same curve since the depolarizing channel is spherically symmetric on the Bloch sphere. Applying the QEC procedure results in a reduced logical error, as expected.  The spinor QEC is generally more effective towards the equator of the Bloch sphere, as predicted by { the arguments in Sec. \ref{sec:ideal}}.  Near the $ z$-poles the code is less effective due to the missing states in the correction operation (\ref{ucdef}).  Each of the curves follow an exponential dependence of the approximate form $ \epsilon_\text{L} \approx  (1 - \exp(-\gamma_\text{L} t ))/2 $, where $ \gamma_\text{L} $ is the logical error rate per cycle.  For many cycles, the states eventually approach $ \epsilon_\text{L} \rightarrow 1/2 $ since the expectation values all approach $ \langle \vec{S} \rangle \rightarrow 0 $.  

Figure \ref{fig4}(b) shows the logical error rate $ \gamma_\text{L} $ as a function of physical error probability for various code sizes. For the case without QEC, all the lines with different $ N $ fall on a single line $ \gamma_L = 4p/3 $. This is because the error channel gives the same effect on all qubits, and spin coherent states are a product state of qubits.   When QEC is included, below the error $ p = 3/4 $ the spinor code shows a consistent improvement with $ N $. All lines pass through the point $ p =3/4 $ and $ \gamma_{\text L} =1 $.  For different initial states, the logical error rates $ \gamma_\text{L} $ change but the crossover point is unchanged.  

{ To understand why the crossover point is at $ p = 0.75 $, note that the depolarizing channel can be equally written as
\begin{align}
    \rho \rightarrow (1- \frac{4p}{3}) \rho + \frac{4p}{3} \frac{I}{2} .
\end{align}
At the  error probability $ p = 3/4 $,} an arbitrary state is mapped to a completely mixed state $ \rho = (I/2)^{\otimes N} $ in a single application of the depolarizing channel. {  At this point all quantum information of the original state is erased and the encoded state is obviously not recoverable.  For error probabilities that are slightly less than this $ p = 3(1- \epsilon)/4 $ for $ \epsilon > 0 $, an initially pure state (\ref{encoding}) becomes
\begin{align}
 \rho =  \left[  \epsilon |\psi_0 \rangle \langle \psi_0 | + (1- \epsilon ) \frac{I}{2}  \right]^{\otimes N}  .
\end{align}
This state has expectation values 
\begin{align}
   \langle S_i \rangle = N \epsilon \langle \psi_0 | \sigma^i  | \psi_0 \rangle  .
\end{align}
Hence for $ N $ large, it is always possible to counteract a small $ \epsilon $ during the decoding and identify the original state.  Thus at a fundamental level, state recovery should always be possible for any $ p < 3/4 $. We note that upper bounds on the highest possible threshold have been estimated to be 25\% based on no-cloning theorem arguments \cite{smith2006upper}.   In our case, since our states are ``pre-cloned'' as shown in Fig. \ref{fig1}(a), the upper bounds do not apply and higher thresholds are obtained. 
}


In the context of quantum fault-tolerance, the crossover of the curves as seen in Fig. \ref{fig4}(b) is considered the hallmark of the error threshold \cite{google2023suppressing,stephens2014fault,fowler2012towards}.  It is tempting to conclude from this that the code-capacity threshold for the spinor code is 75 \%.  However, the crossover criterion generally assumes that as the code size is increased, the logical error rate can be decreased to arbitrarily low levels \cite{Zhao_Liu_2024a}. Physically, this means that the QEC code can asymptotically correct all the error types that it is given. In an approximate QEC code, there is however the possibility that the logical error rates saturate, since it can only partially correct the error that are applied to it. For this reason, it is important to in our case to check whether logical error rates truly go to zero.  In Fig. \ref{fig4}(c),  we plot the logical error rates versus inverse qubit number such that the trend towards $ 1/N \rightarrow 0 $ can be seen.  We conservatively extrapolate with a linear dependence in $ 1/ N $ to determine which physical error $ p $ attain $ \gamma_{\text{L}} = 0$.  This is a conservative dependence since the dependence is concave and it is likely that larger $ p $ also in fact go to $ \gamma_{\text{L}} = 0$ as $ N \rightarrow \infty $.  We observe that for physical errors in the region $ p \lesssim 0.32 $ the logical error rate indeed appears to extrapolate to $ \gamma_{\text{L}} \rightarrow 0 $, confirming that the threshold has been attained.  Hence we estimate the code capacity threshold to be in the region $ 0.32 \lesssim p_\text{th} < 0.75 $. We attribute the reduced threshold to the fact that lower error probabilities are required to make the spinor code a QEC code, as shown in the discussion in Sec. \ref{sec:knilllaflammeconds}.  

In a practical implementation of the spinor code, there are other sources of error, due to imperfections introduced in the measurement procedure \cite{fowler2012surface}.  Specifically, measurement readouts require ancilla qubits which may suffer initialization and measurement errors.  
The logical error rates with these additional imperfections are shown in Figs. \ref{fig4}(d) (see Appendix \ref{app:errsim} for details).  As expected, the additional errors tend the increase the logical errors $ \epsilon_\text{L} $ and its rate $ \gamma_\text{L} $.  The crossover point of the logical error rates remains the same $ p_{\text{th}} = 0.75$.  Again examining the scaling with $ N $ (Fig. \ref{fig4}(e)), and performing a conservative linear extrapolation, we see that error rates $ p \lesssim 0.09 $ are required in order to see a trend towards zero logical error.  Thus in this case we estimate that the phenomenological threshold to be $ 0.09  \lesssim p_\text{th} < 0.75 $.

\section{Conclusions}

In this paper, we introduced and analyzed the spinor code, based on subspaces with fixed total spin quantum number.  One of the unique features of the code is that it is a nonlinear QEC, such that the mapping between the original qubit state and the encoded state is nonlinear. The code does not follow the stabilizer formalism, and thus is also an example of a non-stabilizer code. 
Under a tailored set of errors, the spinor code can also be a standard QEC satisfying Knill-Laflamme conditions.  Under more naturally occurring single qubit Pauli errors, the code approximately satisfies a QEC under certain assumptions.  We showed explicitly that errors map between spaces with small deformation for the encoding that are consider.  Directly simulating the performance of the code, we evaluated code-capacity threshold to be in the range of 32-75 \%, which is one of the highest { code-capacity } thresholds for a QEC code for symmetric Pauli errors, to our knowledge.  When including initialization errors and measurement errors, the  { phenomenological} threshold is estimated to be 9-75 \%.  The large range in the threshold come from the fact that as we deal with an approximate QEC, we are cautious to take at face value the standard crossover test for the threshold.  The lower end of the threshold estimate is obtained by assuming a linear extrapolation in $ 1/N $, which is likely to be overly pessimistic. We emphasize that this scheme is intrinsically nonlinear, and thus the Knill-Laflamme conditions which presuppose a linear subspace structure, are not generally satisfied except under strong constraints on the parameter m. Nevertheless, numerical results demonstrate stronger performance, attributable to both the nonlinear nature of the code and the use of spin expectations as a performance metric.   

The practical experimental implementation of the spinor code will depend heavily upon the particular platform so we leave a detailed investigation of this as future work. Here, we briefly comment on the most important operations.  The key operations that need to be performed are the projection operation (\ref{psldef}) and correction operations (\ref{ucdef}). For a standard gate-based quantum computing architecture, several methods have been developed to implement the projection operation \cite{siwach2021filtering}. 
The correction operation will require a suitable decomposition to a quantum circuit.  Circuit recompilation techniques \cite{jones2022robust,chen2024efficient}, may be one way to find efficient circuits to implement the correction operation.  Overall, the gate-based quantum computing approach may be relatively limited in terms of scalability due to the modest sizes of quantum computers available today.  An alternative is to use atomic gas ensembles, where the hyperfine ground states are decoupled from the motional degrees of freedom.  The physical setup is described in numerous past works, see for example Refs.  \cite{byrnes2024multipartite,byrnes2021quantum,byrnes2015macroscopic}. The potential for scaling here is  enormous as cold atomic ensembles routinely contain $ N = 10^3 $ atoms \cite{riedel2010atom} and $ N = 10^{12} $ for hot ensembles \cite{julsgaard2001experimental}. The primary difficulty with implementing spinor QEC in these systems will be to perform the projection operation and correction operations, which both require a high level of control of the underlying qubits.  Typically in such systems, it is difficult to perform control of the individual qubits, and only globally symmetric operations can be performed, i.e. Hamiltonians involving the total spin operators $ S_j $.  Regarding the correction operation, it is likely that (\ref{ucdef}) is merely one choice of unitary that provides a corrective capability, and others can also equally give an improvement \cite{mao2023measurement}.  It is possible that there are natural processes that provide a natural QEC capability.  As observed experimentally via the Identical Spin Rotation Effect (ISRE) \cite{deutsch2010spin}, a spontaneous rephasing of atomic gas spins may be observed, which may have interpretations of a naturally occurring spinor QEC.  

We have only examined the spinor code for a quantum memory, where the task is to store quantum information.  To be used in the context of quantum computing, one must also consider how the quantum gates are implemented.  The spin coherent state encoding again is highly natural in this context, as single qubit operations are mapped to total spin rotations, which are of the form $ e^{i S_j \theta} $.  For the underlying qubits, this is simply product of single qubit gates, and hence is transversal. It is interesting that the gate is transversal, independent of the rotation angle $ \theta $, hence includes non-Clifford (magic) gates.  Thus producing non-Clifford states in a robust manner should be no impediment for the spinor code.  Due to the nonlinear and nonstabilizer nature of our code, the arguments in Refs. \cite{eastin2009restrictions,zeng2011transversality} do not apply.  What is a more difficult question in our context is the equivalent for the two qubit gate.  Multipartite entangled generalizations of the spinor code can be certainly be defined \cite{byrnes2024multipartite}, however, { entangling gates to manipulate} them are more difficult to define. Whether such entangling operations can be defined transversally we leave as an open question.  

The current QEC procedure is merely one example of a nonlinear QEC and was physically motivated by spinor states \cite{byrnes2024multipartite}.  More generally, it is possible that nonlinear QEC may be a powerful way of attaining high threshold codes. { The reason for the high thresholds that we have observed may be attributed to the high level of duplication that is present in the encoding (\ref{encoding}).  The nonlinear encoding allows it to evade upper bounds in error thresholds set by the no-cloning theorem \cite{smith2006upper}. }
We note that our code has elements of continuous variables quantum information, such as in the use of spin coherent states, which have many  similarities to coherent states \cite{byrnes2021quantum}.  It is well-known that continuous variables has various difficulties in the context of QEC, and one must turn to non-Gaussian states \cite{niset2009no}.
A major difference of our case to the continuous variable QEC scenario is that errors remain discrete even in the limit of large $ N $, as a single qubit error creates a transition between the $ s=N/2 \rightarrow N/2 - 1$ spaces, which always remain orthogonal. Pauli errors also possess a { hierarchal}  structure, where the probability of multiple errors are exponentially suppressed.  In this regard, the spinor code is more similar to qubit-based codes.


\acknowledgments

This work is supported by the SMEC Scientific Research Innovation Project (2023ZKZD55); the National Natural Science Foundation of China (92576102); the Science and Technology Commission of Shanghai Municipality (22ZR1444600); the NYU Shanghai Boost Fund; the China Foreign Experts Program (G2021013002L); the NYU-ECNU Institute of Physics at NYU Shanghai; the NYU Shanghai Major-Grants Seed Fund; and Tamkeen under the NYU Abu Dhabi Research Institute grant CG008.

\bibliography{paper}

@PREAMBLE{
 "\providecommand{\noopsort}[1]{}" 
 # "\providecommand{\singleletter}[1]{#1}%" 
}

@article{bravyi2024high,
  title={High-threshold and low-overhead fault-tolerant quantum memory},
  author={Bravyi, Sergey and Cross, Andrew W and Gambetta, Jay M and Maslov, Dmitri and Rall, Patrick and Yoder, Theodore J},
  journal={Nature},
  volume={627},
  number={8005},
  pages={778--782},
  year={2024},
  publisher={Nature Publishing Group UK London}
}

@article{wu2022erasure,
  title={Erasure conversion for fault-tolerant quantum computing in alkaline earth Rydberg atom arrays},
  author={Wu, Yue and Kolkowitz, Shimon and Puri, Shruti and Thompson, Jeff D},
  journal={Nature communications},
  volume={13},
  number={1},
  pages={4657},
  year={2022},
  publisher={Nature Publishing Group UK London}
}

@article{tuckett2018ultrahigh,
  title={Ultrahigh error threshold for surface codes with biased noise},
  author={Tuckett, David K and Bartlett, Stephen D and Flammia, Steven T},
  journal={Physical review letters},
  volume={120},
  number={5},
  pages={050505},
  year={2018},
  publisher={APS}
}

@article{steane1996multiple,
  title={Multiple-particle interference and quantum error correction},
  author={Steane, Andrew},
  journal={Proceedings of the Royal Society of London. Series A: Mathematical, Physical and Engineering Sciences},
  volume={452},
  number={1954},
  pages={2551--2577},
  year={1996},
  publisher={The Royal Society London}
}

@article{breuckmann2021quantum,
  title={Quantum low-density parity-check codes},
  author={Breuckmann, Nikolas P and Eberhardt, Jens Niklas},
  journal={Prx Quantum},
  volume={2},
  number={4},
  pages={040101},
  year={2021},
  publisher={APS}
}

@book{smith2006upper,
  title={Upper and lower bounds on quantum codes},
  author={Smith, Graeme Stewart Baird},
  year={2006},
  publisher={California Institute of Technology}
}

@article{niset2009no,
  title={No-go theorem for Gaussian quantum error correction},
  author={Niset, Julien and Fiur{\'a}{\v{s}}ek, Jarom{\'\i}r and Cerf, Nicolas J},
  journal={Physical review letters},
  volume={102},
  number={12},
  pages={120501},
  year={2009},
  publisher={APS}
}

@book{byrnes2021quantum,
  title={Quantum atom optics: Theory and applications to quantum technology},
  author={Byrnes, Tim and Ilo-Okeke, Ebubechukwu O},
  year={2021},
  publisher={Cambridge university press}
}

@article{byrnes2012macroscopic,
  title={Macroscopic quantum computation using {Bose}-{Einstein} condensates},
  author={Byrnes, Tim and Wen, Kai and Yamamoto, Yoshihisa},
  journal={Physical Review A},
  volume={85},
  number={4},
  pages={040306(R)},
  year={2012},
  publisher={APS}
}

@article{riedel2010atom,
  title={Atom-chip-based generation of entanglement for quantum metrology},
  author={Riedel, Max F and B{\"o}hi, Pascal and Li, Yun and H{\"a}nsch, Theodor W and Sinatra, Alice and Treutlein, Philipp},
  journal={Nature},
  volume={464},
  number={7292},
  pages={1170--1173},
  year={2010},
  publisher={Nature Publishing Group UK London}
}

@article{byrnes2015macroscopic,
  title={Macroscopic quantum information processing using spin coherent states},
  author={Byrnes, Tim and Rosseau, Daniel and Khosla, Megha and Pyrkov, Alexey and Thomasen, Andreas and Mukai, Tetsuya and Koyama, Shinsuke and Abdelrahman, Ahmed and Ilo-Okeke, Ebubechukwu},
  journal={Optics Communications},
  volume={337},
  pages={102--109},
  year={2015},
  publisher={Elsevier}
}

@article{mohseni2021error,
  title={Error suppression in adiabatic quantum computing with qubit ensembles},
  author={Mohseni, Naeimeh and Narozniak, Marek and Pyrkov, Alexey N and Ivannikov, Valentin and Dowling, Jonathan P and Byrnes, Tim},
  journal={npj Quantum Information},
  volume={7},
  number={1},
  pages={71},
  year={2021},
  publisher={Nature Publishing Group UK London}
}

@article{reichert2022nonlinear,
  title={Nonlinear quantum error correction},
  author={Reichert, Maximilian and Tessler, Louis W and Bergmann, Marcel and van Loock, Peter and Byrnes, Tim},
  journal={Physical Review A},
  volume={105},
  number={6},
  pages={062438},
  year={2022},
  publisher={APS}
}

@article{julsgaard2001experimental,
  title={Experimental long-lived entanglement of two macroscopic objects},
  author={Julsgaard, Brian and Kozhekin, Alexander and Polzik, Eugene S},
  journal={Nature},
  volume={413},
  number={6854},
  pages={400--403},
  year={2001},
  publisher={Nature Publishing Group UK London}
}

@article{gross2012spin,
  title={Spin squeezing, entanglement and quantum metrology with {Bose}--{Einstein} condensates},
  author={Gross, Christian},
  journal={Journal of Physics B: Atomic, Molecular and Optical Physics},
  volume={45},
  number={10},
  pages={103001},
  year={2012},
  publisher={IOP Publishing}
}

@article{deutsch2010spin,
  title={Spin self-rephasing and very long coherence times in a trapped atomic ensemble},
  author={Deutsch, Christian and Ramirez-Martinez, Fernando and Lacro{\^u}te, Clement and Reinhard, Friedemann and Schneider, Tobias and Fuchs, Jean-No{\"e}l and Pi{\'e}chon, Fr{\'e}d{\'e}ric and Lalo{\"e}, Franck and Reichel, Jakob and Rosenbusch, Peter},
  journal={Physical review letters},
  volume={105},
  number={2},
  pages={020401},
  year={2010},
  publisher={APS}
}

@misc{nielsen2002quantum,
  title={Quantum computation and quantum information},
  author={Nielsen, Michael A and Chuang, Isaac},
  year={2002},
  publisher={American Association of Physics Teachers}
}

@inproceedings{grover1996fast,
  title={A fast quantum mechanical algorithm for database search},
  author={Grover, Lov K},
  booktitle={Proceedings of the twenty-eighth annual ACM symposium on Theory of computing},
  pages={212--219},
  year={1996}
}

@inproceedings{shor1994algorithms,
  title={Algorithms for quantum computation: discrete logarithms and factoring},
  author={Shor, Peter W},
  booktitle={Proceedings 35th annual symposium on foundations of computer science},
  pages={124--134},
  year={1994},
  organization={Ieee}
}

@article{cory1998experimental,
  title={Experimental quantum error correction},
  author={Cory, David G and Price, MD and Maas, W and Knill, Emanuel and Laflamme, Raymond and Zurek, Wojciech H and Havel, Timothy F and Somaroo, Shyamal S},
  journal={Physical Review Letters},
  volume={81},
  number={10},
  pages={2152},
  year={1998},
  publisher={APS}
}

@article{pittman2005demonstration,
  title={Demonstration of quantum error correction using linear optics},
  author={Pittman, TB and Jacobs, BC and Franson, JD},
  journal={Physical Review A—Atomic, Molecular, and Optical Physics},
  volume={71},
  number={5},
  pages={052332},
  year={2005},
  publisher={APS}
}

@article{chiaverini2004realization,
  title={Realization of quantum error correction},
  author={Chiaverini, John and Leibfried, Dietrich and Schaetz, Tobias and Barrett, Murray D and Blakestad, RB and Britton, Joseph and Itano, Wayne M and Jost, John D and Knill, Emanuel and Langer, Christopher and others},
  journal={Nature},
  volume={432},
  number={7017},
  pages={602--605},
  year={2004},
  publisher={Nature Publishing Group UK London}
}

@article{schindler2011experimental,
  title={Experimental repetitive quantum error correction},
  author={Schindler, Philipp and Barreiro, Julio T and Monz, Thomas and Nebendahl, Volckmar and Nigg, Daniel and Chwalla, Michael and Hennrich, Markus and Blatt, Rainer},
  journal={Science},
  volume={332},
  number={6033},
  pages={1059--1061},
  year={2011},
  publisher={American Association for the Advancement of Science}
}

@article{reed2012realization,
  title={Realization of three-qubit quantum error correction with superconducting circuits},
  author={Reed, Matthew D and DiCarlo, Leonardo and Nigg, Simon E and Sun, Luyan and Frunzio, Luigi and Girvin, Steven M and Schoelkopf, Robert J},
  journal={Nature},
  volume={482},
  number={7385},
  pages={382--385},
  year={2012},
  publisher={Nature Publishing Group UK London}
}

@book{gottesman1997stabilizer,
  title={Stabilizer codes and quantum error correction},
  author={Gottesman, Daniel},
  year={1997},
  publisher={California Institute of Technology}
}

@article{wootters1982single,
  title={A single quantum cannot be cloned},
  author={Wootters, William K and Zurek, Wojciech H},
  journal={Nature},
  volume={299},
  number={5886},
  pages={802--803},
  year={1982},
  publisher={Nature Publishing Group UK London}
}

@article{byrnes2024multipartite,
  title={Multipartite spin coherent states and spinor states},
  author={Byrnes, Tim},
  journal={Physical Review A},
  volume={109},
  number={2},
  pages={022438},
  year={2024},
  publisher={APS}
}

@article{fowler2012surface,
  title={Surface codes: Towards practical large-scale quantum computation},
  author={Fowler, Austin G and Mariantoni, Matteo and Martinis, John M and Cleland, Andrew N},
  journal={Physical Review A—Atomic, Molecular, and Optical Physics},
  volume={86},
  number={3},
  pages={032324},
  year={2012},
  publisher={APS}
}

@article{fowler2009high,
  title={High-threshold universal quantum computation on the surface code},
  author={Fowler, Austin G and Stephens, Ashley M and Groszkowski, Peter},
  journal={Physical Review A—Atomic, Molecular, and Optical Physics},
  volume={80},
  number={5},
  pages={052312},
  year={2009},
  publisher={APS}
}

@article{acharya2025quantum,
  title={Quantum error correction below the surface code threshold},
  author={{Google Quantum AI and Collaborators}},
  journal={Nature},
  volume={638},
  pages={920-926},
  year={2025},
  publisher={Nature Publishing Group UK London}
}

@article{paetznick2024demonstration,
  title={Demonstration of logical qubits and repeated error correction with better-than-physical error rates},
  author={Paetznick, A and da Silva, MP and Ryan-Anderson, C and Bello-Rivas, JM and Campora III, JP and Chernoguzov, A and Dreiling, JM and Foltz, C and Frachon, F and Gaebler, JP and others},
  journal={arXiv preprint arXiv:2404.02280},
  year={2024}
}

@article{ofek2016extending,
  title={Extending the lifetime of a quantum bit with error correction in superconducting circuits},
  author={Ofek, Nissim and Petrenko, Andrei and Heeres, Reinier and Reinhold, Philip and Leghtas, Zaki and Vlastakis, Brian and Liu, Yehan and Frunzio, Luigi and Girvin, Steven M and Jiang, Liang and others},
  journal={Nature},
  volume={536},
  number={7617},
  pages={441--445},
  year={2016},
  publisher={Nature Publishing Group UK London}
}

@article{ryan2021realization,
  title={Realization of real-time fault-tolerant quantum error correction},
  author={Ryan-Anderson, Ciaran and Bohnet, Justin G and Lee, Kenny and Gresh, Daniel and Hankin, Aaron and Gaebler, JP and Francois, David and Chernoguzov, Alexander and Lucchetti, Dominic and Brown, Natalie C and others},
  journal={Physical Review X},
  volume={11},
  number={4},
  pages={041058},
  year={2021},
  publisher={APS}
}

@article{krinner2022realizing,
  title={Realizing repeated quantum error correction in a distance-three surface code},
  author={Krinner, Sebastian and Lacroix, Nathan and Remm, Ants and Di Paolo, Agustin and Genois, Elie and Leroux, Catherine and Hellings, Christoph and Lazar, Stefania and Swiadek, Francois and Herrmann, Johannes and others},
  journal={Nature},
  volume={605},
  number={7911},
  pages={669--674},
  year={2022},
  publisher={Nature Publishing Group UK London}
}

@article{sivak2023real,
  title={Real-time quantum error correction beyond break-even},
  author={Sivak, Volodymyr V and Eickbusch, Alec and Royer, Baptiste and Singh, Shraddha and Tsioutsios, Ioannis and Ganjam, Suhas and Miano, Alessandro and Brock, BL and Ding, AZ and Frunzio, Luigi and others},
  journal={Nature},
  volume={616},
  number={7955},
  pages={50--55},
  year={2023},
  publisher={Nature Publishing Group UK London}
}

@article{google2023suppressing,
  title={Suppressing quantum errors by scaling a surface code logical qubit},
  author={{Google Quantum AI}},
  journal={Nature},
  volume={614},
  number={7949},
  pages={676--681},
  year={2023},
  publisher={Nature Publishing Group UK London}
}

@article{bluvstein2024logical,
  title={Logical quantum processor based on reconfigurable atom arrays},
  author={Bluvstein, Dolev and Evered, Simon J and Geim, Alexandra A and Li, Sophie H and Zhou, Hengyun and Manovitz, Tom and Ebadi, Sepehr and Cain, Madelyn and Kalinowski, Marcin and Hangleiter, Dominik and others},
  journal={Nature},
  volume={626},
  number={7997},
  pages={58--65},
  year={2024},
  publisher={Nature Publishing Group UK London}
}

@article{gupta2024encoding,
  title={Encoding a magic state with beyond break-even fidelity},
  author={Gupta, Riddhi S and Sundaresan, Neereja and Alexander, Thomas and Wood, Christopher J and Merkel, Seth T and Healy, Michael B and Hillenbrand, Marius and Jochym-O’Connor, Tomas and Wootton, James R and Yoder, Theodore J and others},
  journal={Nature},
  volume={625},
  number={7994},
  pages={259--263},
  year={2024},
  publisher={Nature Publishing Group UK London}
}

@article{livingston2022experimental,
  title={Experimental demonstration of continuous quantum error correction},
  author={Livingston, William P and Blok, Machiel S and Flurin, Emmanuel and Dressel, Justin and Jordan, Andrew N and Siddiqi, Irfan},
  journal={Nature communications},
  volume={13},
  number={1},
  pages={2307},
  year={2022},
  publisher={Nature Publishing Group UK London}
}

@article{bartolucci2023fusion,
  title={Fusion-based quantum computation},
  author={Bartolucci, Sara and Birchall, Patrick and Bombin, Hector and Cable, Hugo and Dawson, Chris and Gimeno-Segovia, Mercedes and Johnston, Eric and Kieling, Konrad and Nickerson, Naomi and Pant, Mihir and others},
  journal={Nature Communications},
  volume={14},
  number={1},
  pages={912},
  year={2023},
  publisher={Nature Publishing Group UK London}
}

@article{bonilla2021xzzx,
  title={The XZZX surface code},
  author={Bonilla Ataides, J Pablo and Tuckett, David K and Bartlett, Stephen D and Flammia, Steven T and Brown, Benjamin J},
  journal={Nature communications},
  volume={12},
  number={1},
  pages={2172},
  year={2021},
  publisher={Nature Publishing Group UK London}
}

@inproceedings{shor1996fault,
  title={Fault-tolerant quantum computation},
  author={Shor, Peter W},
  booktitle={Proceedings of 37th conference on foundations of computer science},
  pages={56--65},
  year={1996},
  organization={IEEE}
}

@inproceedings{aharonov1997fault,
  title={Fault-tolerant quantum computation with constant error},
  author={Aharonov, Dorit and Ben-Or, Michael},
  booktitle={Proceedings of the twenty-ninth annual ACM symposium on Theory of computing},
  pages={176--188},
  year={1997}
}

@article{knill1998resilient,
  title={Resilient quantum computation},
  author={Knill, Emanuel and Laflamme, Raymond and Zurek, Wojciech H},
  journal={Science},
  volume={279},
  number={5349},
  pages={342--345},
  year={1998},
  publisher={American Association for the Advancement of Science}
}

@article{kitaev2003fault,
  title={Fault-tolerant quantum computation by anyons},
  author={Kitaev, A Yu},
  journal={Annals of physics},
  volume={303},
  number={1},
  pages={2--30},
  year={2003},
  publisher={Elsevier}
}

@article{shor1995scheme,
  title={Scheme for reducing decoherence in quantum computer memory},
  author={Shor, Peter W},
  journal={Physical review A},
  volume={52},
  number={4},
  pages={R2493},
  year={1995},
  publisher={APS}
}

@article{peres1985reversible,
  title={Reversible logic and quantum computers},
  author={Peres, Asher},
  journal={Physical review A},
  volume={32},
  number={6},
  pages={3266},
  year={1985},
  publisher={APS}
}

@article{devitt2013quantum,
  title={Quantum error correction for beginners},
  author={Devitt, Simon J and Munro, William J and Nemoto, Kae},
  journal={Reports on Progress in Physics},
  volume={76},
  number={7},
  pages={076001},
  year={2013},
  publisher={IOP Publishing}
}

@article{roffe2019quantum,
  title={Quantum error correction: an introductory guide},
  author={Roffe, Joschka},
  journal={Contemporary Physics},
  volume={60},
  number={3},
  pages={226--245},
  year={2019},
  publisher={Taylor \& Francis}
}

@article{terhal2015quantum,
  title={Quantum error correction for quantum memories},
  author={Terhal, Barbara M},
  journal={Reviews of Modern Physics},
  volume={87},
  number={2},
  pages={307--346},
  year={2015},
  publisher={APS}
}

@article{steane2006tutorial,
  title={A tutorial on quantum error correction},
  author={Steane, Andrew M},
  journal={Quantum Computers, Algorithms and Chaos},
  pages={1--32},
  year={2006},
  publisher={IOS Press}
}

@article{dennis2002topological,
  title={Topological quantum memory},
  author={Dennis, Eric and Kitaev, Alexei and Landahl, Andrew and Preskill, John},
  journal={Journal of Mathematical Physics},
  volume={43},
  number={9},
  pages={4452--4505},
  year={2002},
  publisher={American Institute of Physics}
}

@article{raussendorf2007fault,
  title={Fault-tolerant quantum computation with high threshold in two dimensions},
  author={Raussendorf, Robert and Harrington, Jim},
  journal={Physical review letters},
  volume={98},
  number={19},
  pages={190504},
  year={2007},
  publisher={APS}
}

@article{knill1997theory,
  title={Theory of quantum error-correcting codes},
  author={Knill, Emanuel and Laflamme, Raymond},
  journal={Physical Review A},
  volume={55},
  number={2},
  pages={900},
  year={1997},
  publisher={APS}
}

@article{siwach2021filtering,
  title={Filtering states with total spin on a quantum computer},
  author={Siwach, Pooja and Lacroix, Denis},
  journal={Physical Review A},
  volume={104},
  number={6},
  pages={062435},
  year={2021},
  publisher={APS}
}

@article{jones2022robust,
  title={Robust quantum compilation and circuit optimisation via energy minimisation},
  author={Jones, Tyson and Benjamin, Simon C},
  journal={Quantum},
  volume={6},
  pages={628},
  year={2022},
  publisher={Verein zur F{\"o}rderung des Open Access Publizierens in den Quantenwissenschaften}
}

@article{chen2024efficient,
  title={Efficient preparation of the AKLT state with measurement-based imaginary time evolution},
  author={Chen, Tianqi and Byrnes, Tim},
  journal={Quantum},
  volume={8},
  pages={1557},
  year={2024},
  publisher={Verein zur F{\"o}rderung des Open Access Publizierens in den Quantenwissenschaften}
}

@book{gerry2023introductory,
  title={Introductory quantum optics},
  author={Gerry, Christopher C and Knight, Peter L},
  year={2023},
  publisher={Cambridge university press}
}

@article{stephens2014fault,
  title={Fault-tolerant thresholds for quantum error correction with the surface code},
  author={Stephens, Ashley M},
  journal={Physical Review A},
  volume={89},
  number={2},
  pages={022321},
  year={2014},
  publisher={APS}
}

@article{fowler2012towards,
  title={Towards practical classical processing for the surface code},
  author={Fowler, Austin G and Whiteside, Adam C and Hollenberg, Lloyd CL},
  journal={Physical review letters},
  volume={108},
  number={18},
  pages={180501},
  year={2012},
  publisher={APS}
}

@article{higgott2021subsystem,
  title={Subsystem codes with high thresholds by gauge fixing and reduced qubit overhead},
  author={Higgott, Oscar and Breuckmann, Nikolas P},
  journal={Physical Review X},
  volume={11},
  number={3},
  pages={031039},
  year={2021},
  publisher={APS}
}

@article{mao2023measurement,
  title={Measurement-based deterministic imaginary time evolution},
  author={Mao, Yuping and Chaudhary, Manish and Kondappan, Manikandan and Shi, Junheng and Ilo-Okeke, Ebubechukwu O and Ivannikov, Valentin and Byrnes, Tim},
  journal={Physical Review Letters},
  volume={131},
  number={11},
  pages={110602},
  year={2023},
  publisher={APS}
}

@article{bravyi2012subsystem,
  title={Subsystem surface codes with three-qubit check operators},
  author={Bravyi, Sergey and Duclos-Cianci, Guillaume and Poulin, David and Suchara, Martin},
  journal={arXiv preprint arXiv:1207.1443},
  year={2012}
}

@article{Zhao_Liu_2024a, 
    title={Extracting error thresholds through the framework of approximate quantum error correction condition}, 
    volume={6}, 
    number={4}, 
    journal={Physical Review Research}, 
    author={Zhao, Yuanchen and Liu, Dong E.}, 
    year={2024}, 
    month={Dec}
}

@article{eastin2009restrictions,
  title={Restrictions on transversal encoded quantum gate sets},
  author={Eastin, Bryan and Knill, Emanuel},
  journal={Physical review letters},
  volume={102},
  number={11},
  pages={110502},
  year={2009},
  publisher={APS}
}

@article{zeng2011transversality,
  title={Transversality versus universality for additive quantum codes},
  author={Zeng, Bei and Cross, Andrew and Chuang, Isaac L},
  journal={IEEE Transactions on Information Theory},
  volume={57},
  number={9},
  pages={6272--6284},
  year={2011},
  publisher={IEEE}
}

@article{omanakuttan2024fault,
  title={Fault-tolerant quantum computation using large spin-cat codes},
  author={Omanakuttan, Sivaprasad and Buchemmavari, Vikas and Gross, Jonathan A and Deutsch, Ivan H and Marvian, Milad},
  journal={PRX Quantum},
  volume={5},
  number={2},
  pages={020355},
  year={2024},
  publisher={APS}
}

@inproceedings{omanakuttan2023spin,
  title={Spin squeezed GKP codes for quantum error correction in atomic ensembles},
  author={Omanakuttan, Sivaprasad and Volkoff, Tyler},
  booktitle={APS Division of Atomic, Molecular and Optical Physics Meeting Abstracts},
  volume={2023},
  pages={F01--054},
  year={2023}
}

@article{brion2008error,
  title={Error correction in ensemble registers for quantum repeaters and quantum computers},
  author={Brion, Etienne and Pedersen, Line Hjortsh{\o}j and Saffman, Mark and M{\o}lmer, Klaus},
  journal={Physical review letters},
  volume={100},
  number={11},
  pages={110506},
  year={2008},
  publisher={APS}
}

@article{ruskai1999pauli,
  title={Pauli exchange errors in quantum computation},
  author={Ruskai, Mary Beth},
  journal={arXiv preprint quant-ph/9906114},
  year={1999}
}

@article{pollatsek2004permutationally,
  title={Permutationally invariant codes for quantum error correction},
  author={Pollatsek, Harriet and Ruskai, Mary Beth},
  journal={Linear algebra and its applications},
  volume={392},
  pages={255--288},
  year={2004},
  publisher={Elsevier}
}

@article{ouyang2014permutation,
  title={Permutation-invariant quantum codes},
  author={Ouyang, Yingkai},
  journal={Physical Review A},
  volume={90},
  number={6},
  pages={062317},
  year={2014},
  publisher={APS}
}

@article{brandao2019quantum,
  title={Quantum error correcting codes in eigenstates of translation-invariant spin chains},
  author={Brandao, Fernando GSL and Crosson, Elizabeth and {\c{S}}ahino{\u{g}}lu, M Burak and Bowen, John},
  journal={Physical review letters},
  volume={123},
  number={11},
  pages={110502},
  year={2019},
  publisher={APS}
}

@article{calderbank1996good,
  title={Good quantum error-correcting codes exist},
  author={Calderbank, A Robert and Shor, Peter W},
  journal={Physical Review A},
  volume={54},
  number={2},
  pages={1098},
  year={1996},
  publisher={APS}
}

@article{steane1996simple,
  title={Simple quantum error-correcting codes},
  author={Steane, Andrew M},
  journal={Physical Review A},
  volume={54},
  number={6},
  pages={4741},
  year={1996},
  publisher={APS}
}

@article{rains1997nonadditive,
  title={A nonadditive quantum code},
  author={Rains, Eric M and Hardin, RH and Shor, Peter W and Sloane, Neil JA},
  journal={Physical Review Letters},
  volume={79},
  number={5},
  pages={953},
  year={1997},
  publisher={APS}
}

@inproceedings{cross2008codeword,
  title={Codeword stabilized quantum codes},
  author={Cross, Andrew and Smith, Graeme and Smolin, John A and Zeng, Bei},
  booktitle={2008 IEEE International Symposium on Information Theory},
  pages={364--368},
  year={2008},
  organization={IEEE}
}

@article{chuang2009codeword,
  title={Codeword stabilized quantum codes: Algorithm and structure},
  author={Chuang, Isaac and Cross, Andrew and Smith, Graeme and Smolin, John and Zeng, Bei},
  journal={Journal of Mathematical Physics},
  volume={50},
  number={4},
  year={2009},
  publisher={AIP Publishing}
}

@article{grassl2009generalized,
  title={Generalized concatenated quantum codes},
  author={Grassl, Markus and Shor, Peter and Smith, Graeme and Smolin, John and Zeng, Bei},
  journal={Physical Review A—Atomic, Molecular, and Optical Physics},
  volume={79},
  number={5},
  pages={050306},
  year={2009},
  publisher={APS}
}

@article{smolin2007simple,
  title={Simple family of nonadditive quantum codes},
  author={Smolin, John A and Smith, Graeme and Wehner, Stephanie},
  journal={Physical review letters},
  volume={99},
  number={13},
  pages={130505},
  year={2007},
  publisher={APS}
}

@article{ouyang2017permutation,
  title={Permutation-invariant qudit codes from polynomials},
  author={Ouyang, Yingkai},
  journal={Linear Algebra and its Applications},
  volume={532},
  pages={43--59},
  year={2017},
  publisher={Elsevier}
}

@article{ouyang2019permutation,
  title={Permutation-invariant constant-excitation quantum codes for amplitude damping},
  author={Ouyang, Yingkai and Chao, Rui},
  journal={IEEE Transactions on Information Theory},
  volume={66},
  number={5},
  pages={2921--2933},
  year={2019},
  publisher={IEEE}
}

@article{aydin2024family,
  title={A family of permutationally invariant quantum codes},
  author={Aydin, Arda and Alekseyev, Max A and Barg, Alexander},
  journal={Quantum},
  volume={8},
  pages={1321},
  year={2024},
  publisher={Verein zur F{\"o}rderung des Open Access Publizierens in den Quantenwissenschaften}
}

@article{aydin2026quantum,
  title={Quantum error correction beyond SU (2): Spin, bosonic, and permutation-invariant codes from convex geometry},
  author={Aydin, Arda and Albert, Victor V and Barg, Alexander},
  journal={PRX Quantum},
  volume={7},
  number={1},
  pages={010341},
  year={2026},
  publisher={APS}
}

@article{ouyang2025measurement,
  title={Measurement-free code-switching protocol for low-overhead quantum computation using permutation-invariant codes},
  author={Ouyang, Yingkai and Jing, Yumang and Brennen, Gavin K},
  journal={PRX Quantum},
  volume={6},
  number={4},
  pages={040341},
  year={2025},
  publisher={APS}
}

@article{kubischta2023family,
  title={Family of quantum codes with exotic transversal gates},
  author={Kubischta, Eric and Teixeira, Ian},
  journal={Physical Review Letters},
  volume={131},
  number={24},
  pages={240601},
  year={2023},
  publisher={APS}
}

@article{mirrahimi2014dynamically,
  title={Dynamically protected cat-qubits: a new paradigm for universal quantum computation},
  author={Mirrahimi, Mazyar and Leghtas, Zaki and Albert, Victor V and Touzard, Steven and Schoelkopf, Robert J and Jiang, Liang and Devoret, Michel H},
  journal={New Journal of Physics},
  volume={16},
  number={4},
  pages={045014},
  year={2014},
  publisher={IOP Publishing}
}

@article{michael2016new,
  title={New class of quantum error-correcting codes for a bosonic mode},
  author={Michael, Marios H and Silveri, Matti and Brierley, RT and Albert, Victor V and Salmilehto, Juha and Jiang, Liang and Girvin, Steven M},
  journal={Physical Review X},
  volume={6},
  number={3},
  pages={031006},
  year={2016},
  publisher={APS}
}

@article{PhysRevA.64.012310,
  title = {Encoding a qubit in an oscillator},
  author = {Gottesman, Daniel and Kitaev, Alexei and Preskill, John},
  journal = {Phys. Rev. A},
  volume = {64},
  issue = {1},
  pages = {012310},
  numpages = {21},
  year = {2001},
  month = {Jun},
  publisher = {American Physical Society},
  doi = {10.1103/PhysRevA.64.012310},
  url = {https://link.aps.org/doi/10.1103/PhysRevA.64.012310}
}

@article{bombin2012strong,
  title={Strong resilience of topological codes to depolarization},
  author={Bombin, H{\'e}ctor and Andrist, Ruben S and Ohzeki, Masayuki and Katzgraber, Helmut G and Martin-Delgado, Miguel A},
  journal={Physical Review X},
  volume={2},
  number={2},
  pages={021004},
  year={2012},
  publisher={APS}
}

@article{wootton2012high,
  title={High threshold error correction for the surface code},
  author={Wootton, James R and Loss, Daniel},
  journal={Physical review letters},
  volume={109},
  number={16},
  pages={160503},
  year={2012},
  publisher={APS}
}

@article{wang2011threshold,
author = {Wang, D. S. and Fowler, A. G. and Stephens, A. M. and Hollenberg, L. C. L.},
title = {Threshold error rates for the toric and planar codes},
year = {2010},
issue_date = {May 2010},
publisher = {Rinton Press, Incorporated},
address = {Paramus, NJ},
volume = {10},
number = {5},
journal = {Quantum Info. Comput.},
month = may,
pages = {456–469},
numpages = {14}
}

@article{kuo2026degenerate,
  title={Degenerate quantum erasure decoding},
  author={Kuo, Kao-Yueh and Ouyang, Yingkai},
  journal={npj Quantum Information},
  volume={12},
  number={1},
  pages={75},
  year={2026},
  publisher={Nature Publishing Group UK London}
}

@article{PhysRevResearch.2.043423,
  title = {Decoding across the quantum low-density parity-check code landscape},
  author = {Roffe, Joschka and White, David R. and Burton, Simon and Campbell, Earl},
  journal = {Phys. Rev. Res.},
  volume = {2},
  issue = {4},
  pages = {043423},
  numpages = {13},
  year = {2020},
  month = {Dec},
  publisher = {American Physical Society},
  doi = {10.1103/PhysRevResearch.2.043423},
  url = {https://link.aps.org/doi/10.1103/PhysRevResearch.2.043423}
}

@article{ouyang2026theory,
  title={A theory of quantum error correction for permutation-invariant codes},
  author={Ouyang, Yingkai and Brennen, Gavin K},
  journal={arXiv preprint arXiv:2602.13638},
  year={2026}
}
 

\appendix

\section{Applications of the spinor code}
\label{app:spinorqc}

In this section, we discuss the situations where spinor encoding (\ref{encoding}) is applicable.  

One of the prime applications of the spinor code is spinor quantum computation, which is a framework where qubit ensembles are used to store the quantum information \cite{byrnes2012macroscopic,byrnes2021quantum}.  The scheme shares similarities with the scheme described in Fig. \ref{fig1}(a) in that a quantum computation is implemented with qubit ensembles. In both cases, there is an $ N $-fold duplication of the state of an $ M $-qubit quantum computer.  The main difference between the two schemes is that in spinor quantum computation, the state of the quantum register at any point in time is a multipartite spinor state, whereas in Fig. \ref{fig1}(a), it is a multipartite spin coherent state.  Here, the multipartite spinor state \cite{byrnes2024multipartite} can be written as
\begin{align}
|\Psi \rangle \rangle = \frac{1}{\sqrt{{\cal N}_{\Psi} }}
\left(
\sum_{l_1=0}^1 \dots \sum_{l_M=0}^1 \Psi_{l_1 \dots l_M} 
a_{1,l_1}^\dagger \dots a_{M,l_M}^\dagger 
\right)^N | \text{vac} \rangle
\label{spinorpsi}
\end{align}
where $ a_{m,l}^\dagger $ is a bosonic creation operator for the $m$th ensemble of the $ l$th type, $ | \text{vac} \rangle $ is the bosonic vacuum state, ${\cal N}_{\Psi} $ is a normalization factor, and $  \Psi_{l_1 \dots l_M} $ are the coefficients of a $ M $-qubit wavefunction.   For an encoding of qubits such as that considered in (\ref{encoding}), there are two boson types $ l \in \{0,1 \} $.  Meanwhile, the multipartite spin coherent state is defined as
\begin{align}
| \Psi \rangle^{\otimes N} = \left(
\sum_{l_1=0}^1 \dots \sum_{l_M=0}^1 \Psi_{l_1 \dots l_M} |l_1 \dots l_M  \rangle
\right)^{\otimes N}
\label{scspsi}
\end{align}
where $ |l_1 \dots l_M  \rangle $ is a $ M $-qubit register in the computational basis.  In both cases, the quantum state of an $ M $-qubit register is encoded, hence both are capable of storing the same information, and both are capable of universal quantum computation.  

As shown in Ref. \cite{byrnes2024multipartite}, the spinor states and spin coherent states are equivalent in the case of a single ensemble, but with multiple ensembles, they no longer coincide.  
The main difference between the two classes of states can be understood by examining  the total spin of the states that are present on each ensemble.  For spinor states (\ref{spinorpsi}), when transformed to the language of total spin eigenstates \cite{byrnes2024multipartite}, they can be written entirely in terms of the maximal spin states due to the equivalence
\begin{align}
    \frac{(a^\dagger)^k (b^\dagger)^{N-k}}{\sqrt{k! (N-k)!}} | \text{vac} \rangle \leftrightarrow | s=\frac{N}{2}, 1, k-\frac{N}{2} \rangle   .
\end{align}
This arises because the bosonic form guarantees symmetry under particle interchange, a property shared by the maximal spin sector.  Meanwhile, multipartite spin coherent states may involve non-maximal total spin, i.e. $ s < N/2$.  The consequence of this is that the spinor code is more applicable to spinor states, since the error-free states always have maximal total spin on each ensemble.  The spinor code then acts to return the state on each ensemble back to its maximal spin. However, for entangled multipartite spin coherent states, since the error-free states do not have maximal spin, the spinor QEC will act to modify the state such that each ensemble has the maximal total spin.  The exception to this is when the state (\ref{scspsi}) is a product state with respect to the $ M $ ensembles.  In this case, the spin coherent state and spinor states are equivalent, and the spinor QEC can be applied. 

In addition to the quantum computing scenario, where a quantum circuit takes the state from a known state to an unknown one as in Fig. \ref{fig1}(a), there are other situations where the state is known but requires protection from decoherence.  For example, in quantum metrology applications, spin squeezed states are of particular interest \cite{gross2012spin,riedel2010atom}.  Such states take the form 
\begin{align}
| \xi \rangle & = e^{-i S_z^2 \xi } | \alpha ,  \beta  \rangle \rangle \nonumber \\
& = \sum_{k=0}^N \sqrt{N \choose k} \alpha^k \beta^{N-k} e^{i\xi (k-N/2)^2} |\frac{N}{2}, 1, k-\frac{N}{2} \rangle ,
\end{align}
where a typical choice is $ \alpha = \beta = 1/\sqrt{2} $. The key things to note here are that only maximal total spin states are involved, which is a prerequisite for the use of the spinor code.  Second, the amplitude of the states is no different to (\ref{scsexp}), the only difference is the phase factor  $ e^{i\xi (k-N/2)^2}$.  Thus the arguments relating to the deformation of the state under single qubit Pauli errors hold, where the states have a Gaussian distribution as shown in Fig. \ref{fig2}.  Thus spin squeezed states are another potential application of the spinor code.

\section{Total spin eigenstates}
\label{app:totalspin}

In this section we discuss the definitions of the total spin eigenstates $ |s, l, m \rangle $.  For the eigenvalues $ s, m $, these are defined by (\ref{s2eigs}) and (\ref{szeigs}) respectively.  The $ l $ eigenvalue is not well defined since it is degenerate with respect to both of the operators $ S^2, S_z  $.  Simply finding the mutual eigenstates of these two operators will result in an inconsistent definition of $ l $.  

We define a consistent definition of $ l $ by the following procedure.  For each spin sector $ S $, 
\begin{enumerate}
\item Diagonalize $ S^2 - S_z $ and obtain the eigenstates with maximal $ S_z $ eigenvalue, i.e. $ |s, l, m=s \rangle $. \\
\item If the $ |s, l, s \rangle $ are not mutually orthogonal with respect to the $ l $-label, orthogonalize using a suitable procedure such as Gram-Schmidt orthogonalization such that 
$ |\langle s, l, s | s, l', s \rangle|^2 = \delta_{l l'} $. \\
\item For all $ m \in [-s, s] $ and $ l \in [1, L_s ] $, apply ladder operators and define
\begin{align}
   |s, l, m  \rangle = \frac{(S_{-})^{s-m} |s,l,s \rangle}{\sqrt{\langle s,l,s | (S_{+})^{s-m} (S_{-})^{s-m} |s,l,s \rangle }} 
\end{align}
where $ S_{\pm} = S_x \pm  i S_y $. 
\end{enumerate}
Repeating the above procedure for all spin sectors $ s \in \{N/2, N/2-1, .... \}$ we obtain the full set of total spin eigenstates.  

The degeneracy of the spin sectors is given by 
\begin{align}
L_s = {N \choose N/2-s }- {N \choose N/2-s - 1 }
\end{align}
for $ s \in \{ N/2, N/2-1, \dots \}$, which can be deduced from the number of orthogonal states with maximal $ S_z $ eigenvalue $ |s, l, s \rangle $.  In the context of spinor QEC, the most important sectors are $ s= N/2 $ which has no degeneracy $ L_s = 1 $ and the single error space $ s = N/2 - 1$ with degeneracy $ L_s = N - 1$.

\section{Logical error}
\label{app:logicalerror}

{
In this section, we further discuss why it is important to evaluate the error  after performing the decoding process, and not on the encoded states themselves.  }

When a nonlinear encoding is used such as (\ref{encoding}), quantities such as fidelity can involve a dependence on the size of the code ($N$ in our case) that is not related to the underlying quantum information that is being stored.  For example, compare two encoded qubit states on the equator of the Bloch sphere, separated by an angle $ \delta $.  We may consider the state $ | \frac{1}{\sqrt{2}}, \frac{1}{\sqrt{2}} \rangle \rangle $ to be the original state and the state $ | \frac{1}{\sqrt{2}}, \frac{e^{i \delta  }}{\sqrt{2}} \rangle \rangle $ to be the a state with a logical error.  Evaluating the fidelity of these two states we find
\begin{align}
F_{\text{NL}} & = \Big| \langle \langle \frac{1}{\sqrt{2}}, \frac{1}{\sqrt{2}} | \frac{1}{\sqrt{2}}, \frac{e^{i \delta  }}{\sqrt{2}} \rangle \rangle \Big|^2 \nonumber \\
& = \cos^{2N} \frac{\delta}{2} \approx \exp (- \frac{N \delta^2}{4} ) .
\end{align}
We see that as the code size is increased, the fidelity exponentially decreases, despite the logical information stored in the encoding being the same for all $ N $.  We note that this is not related any approximations of the spinor QEC procedure, it arises simply due to the nonlinear encoding of the quantum information.  

This dependence on the code size does not happen in standard QEC approaches due to the linearity of the encoding. For example, consider a code with logical states $ |0_\text{L} \rangle^{(N)},|1_\text{L} \rangle^{(N)} $.  We have labeled these with a superscript $ N $ to indicate that these states involve $ N $ qubits, which is the code size.    Evaluating the fidelity for the same example, we have
\begin{align}
F_{\text{lin}} & = \Big| ( \frac{1}{\sqrt{2}} \langle 0_\text{L}^{(N)}  | + \frac{1}{\sqrt{2}} \langle 1_\text{L}^{(N)}  | ) ( \frac{1}{\sqrt{2}} | 0_\text{L}^{(N)}  \rangle  + \frac{e^{i \delta  }}{\sqrt{2}} | 1_\text{L}^{(N)}  \rangle ) \Big|^2 \nonumber \\
& = \cos^2 \frac{\delta}{2} \label{linearfid}
\end{align}
which clearly has no dependence on the number of qubits $ N $. 

{
By performing the decoding process before evaluating the error (or fidelity),} it only compares the logical information stored in the encoding (\ref{encoding}).  Evaluating the logical error we have
\begin{align}
    \epsilon_{\text{L}} & =|\sin\frac{\delta}{2}| \\
    & = \sqrt{1-F_{\text{lin}} } ,
\end{align}
which does not have any $ N $-dependence and can be related to (\ref{linearfid}) using standard relations between trace distance and fidelity of a qubit \cite{nielsen2002quantum}.

\section{Effect of Pauli errors on spinor code}
\label{app:deformation}

{
In this section, we evaluate the effect of Pauli errors on the spinor code. We consider applying a phase flip error $ \sigma^z_n $ on the spin coherent state (\ref{scsexp}), then projecting with (\ref{psldef}).  Following (\ref{deformz}),  we have
\begin{align}
|\tilde{\psi} \rangle & = P_{sl} \sigma^z_n |\alpha, \beta \rangle \rangle  \nonumber \\
& = \sum_{k=0}^N \sqrt{N \choose k} \alpha^k \beta^{N-k} D_{sl}^{(n)} (k-\frac{N}{2} ) | \frac{N}{2}, 1, k- \frac{N}{2} \rangle ,
\label{psitildeN2}
\end{align}
where the deformation factors (\ref{dfactors}) are used. This state is then decoded using (\ref{decodedstate}).  The expectation values of the decoded state is
\begin{align}
\langle \sigma^j \rangle = \text{Tr} [ {\cal D} (\rho) \sigma^j ] = \frac{\langle S_j \rangle}{N/2} .
\end{align}
We evaluate the expectation values as
\begin{align}
    \langle S_j \rangle = \frac{\langle \tilde{\psi} | S_j | \tilde{\psi} \rangle}{\langle \tilde{\psi} | \tilde{\psi} \rangle}
    \label{sexp}
\end{align}
due to the unnormalized nature of the state (\ref{psitildeN2}).  For $ s = N/2 $ we are able to obtain analytical expressions for the expectation values and we obtain
\begin{align}
\langle \sigma^x \rangle  & = \frac{N-2}{N} \sin \theta \cos \phi \\
\langle \sigma^y \rangle  & = \frac{N-2}{N} \sin \theta \sin \phi \\
\langle \sigma^z \rangle  & = \frac{\cos \theta ( -2 + 3N + N^2 + (2-3N+N^2) \cos 2 \theta)}{N(1+N + (N-1) \cos 2 \theta)}  \\
p_{sl} & = \langle \tilde{\psi} | \tilde{\psi} \rangle = \frac{N+1 +(N-1)\cos \theta}{N}
\end{align}

For $ s = N/2 -1 $, we perform the same procedure but evaluate the expressions numerically.  Since the only $ l $-dependence of the state (\ref{psitildeN2}) occurs in the normalization of the deformation factor in (\ref{dfactors}), these cancel in (\ref{sexp}) and eventually give no $ l $-dependence.  
}

\section{Approximate Knill-Laflamme conditions for dephasing}

\label{app:klcond}

In this section we show that under the constraint (\ref{klcriterion}) the error spaces are $ m $-independent.  The (unnormalized) eigenvectors of the $ w $ matrix (\ref{wmatrix}) are given by 
\begin{align}
\vec{u}_0 & = ( \frac{\Gamma +\Delta_m }{2a_m} , 1)^T \nonumber \\
\vec{u}_1 & = (\frac{\Gamma-\Delta_m }{2a_m} , 1)^T \nonumber \\
\label{weigenvectors}
\end{align}
The corresponding eigenvalues are 
\begin{align}
\lambda_0 & = \frac{1+  \Delta_m}{2}  \\
\lambda_1 & = \frac{1 -  \Delta_m }{2}  
\label{lambdavals}
\end{align}
where we defined
\begin{align}
    \Delta_m & = \sqrt{4 a_m^2 + (c-d)^2 }  \label{deltadef} \\
    \Gamma & = 1- 2p
\end{align}

Let us now examine the $ m $-dependence of the eigenvectors and eigenvalues.  In the limit that 
\begin{align}
    4 a_m^2 \ll \Gamma^2 ,
    \label{condkl}
\end{align}
we can expand the square root in (\ref{deltadef}) to give
\begin{align}
    \Delta_m \approx \Gamma + \frac{2Na_m^2}{\Gamma} .
\end{align}
In this limit the  eigenvectors take the form 
\begin{align}
\vec{u}_0 & \approx  (1 , 0)^T \nonumber \\
\vec{u}_1 & \approx(0 , 1)^T 
\label{weigenvectors2}
\end{align}
with eigenvalues $ \lambda_0 \approx  (1+ \Gamma)/2 $,  $ \lambda_1 \approx  (1 - \Gamma)/2 $.  These are explicitly independent of $ m $.  
The regime for this is obtained by substituting explicit expressions into (\ref{condkl}), which gives
\begin{align}
4 (m/N)^2 (1-p ) p  & \ll (1-2 p)^2  .
\end{align}
{ 
Using the fact that $ (m/N)^2 \le 1/4 $,  simplification of the above condition gives (\ref{klcriterion}).  }

{
We now also directly quantify the
variation of the $ w $-matrix itself. Define
\begin{equation}
q=\sqrt{p(1-p)},
\end{equation}
such that
\begin{equation}
w(m)
=
\begin{pmatrix}
1-p & q m/N\\
q m/N & p
\end{pmatrix}.
\label{eq:w-matrix-band}
\end{equation}

Consider a band centered at \(m_0=cN\), where \(c\) is independent
of \(N\), with
\begin{equation}
\mathcal{B}_{N}(m_0,K)
=
\left\{
m:\left|m-m_0\right|\leq K\sqrt{N}
\right\},
\end{equation}
where \(K\) is also independent of \(N\). For any \(m\) in this
band, we have
\begin{equation}
w(m)-w(m_0)
=
q\frac{m-m_0}{N}
\begin{pmatrix}
0&1\\
1&0
\end{pmatrix}.
\end{equation}
Since the operator norm of the Pauli matrix appearing above is one,
it follows that
\begin{equation}
\left\|w(m)-w(m_0)\right\|
=
q\frac{|m-m_0|}{N}
\leq
\frac{K\sqrt{p(1-p)}}{\sqrt{N}}.
\label{eq:w-matrix-variation}
\end{equation}
Consequently,
\begin{equation}
w(m)
=
w(m_0)+O\left(N^{-1/2}\right)
\end{equation}
uniformly throughout the band. Equivalently, for any
\(m,n\in\mathcal{B}_{N}(m_0,K)\),
\begin{equation}
\left\|w(m)-w(n)\right\|
\leq
\frac{2K\sqrt{p(1-p)}}{\sqrt{N}}.
\end{equation}

The Knill--Laflamme matrix elements can therefore be written as
\begin{equation}
\langle C_m|E_j^\dagger E_{j'}|C_{m'}\rangle
=
\left[
w_{jj'}(m_0)+r_{jj'}(m)
\right]\delta_{mm'},
\end{equation}
where the remainder satisfies
\begin{equation}
\|r(m)\|
\leq
\frac{K\sqrt{p(1-p)}}{\sqrt{N}}.
\end{equation}
Thus, within any \(O(\sqrt{N})\) band centered at a fixed value of
\(m/N\), the Knill--Laflamme matrix becomes asymptotically constant
at least as fast as \(N^{-1/2}\).

{
\section{Approximate Knill-Laflamme conditions for more general errors}
\label{app:new_akl}
In this section, we generalize the approximate
Knill--Laflamme analysis to more general errors.  We again consider  states whose magnetic quantum-number
distribution is centered at
\begin{equation}
    \frac{m}{N}\approx c,
\end{equation}
where \(c\) is a fixed constant that need not vanish.  A spin
coherent state has a distribution in \(m\) whose center is of order
\(N\), while its width is of order \(\sqrt{N}\).  It is therefore
natural to consider an \(O(\sqrt{N})\) band centered at \(cN\).

\subsection{General band theorem}

Fix a constant
\begin{equation}
    -\frac{1}{2}<c<\frac{1}{2}
\end{equation}
and a constant \(K>0\), both independent of \(N\).  Let
\(\bar m_N(c)\) denote an allowed magnetic quantum number closest to
\(cN\), and define
\begin{equation}
    B_N(c,K)
    :=
    \left\{
    m:
    |m-\bar m_N(c)|\leq K\sqrt{N}
    \right\}.
    \label{eq:moving-band}
\end{equation}
We define the error-free code space basis state:
\begin{equation}
    |C_m\rangle
    :=
    \left|\frac{N}{2},1,m\right\rangle
\end{equation}
and the band projector
\begin{equation}
    P_{N,c}
    :=
    \sum_{m\in B_N(c,K)}
    |C_m\rangle\langle C_m|.
    \label{eq:moving-band-projector}
\end{equation}

Let
\begin{equation}
    \mathcal E=\{E_1,\ldots,E_r\}
\end{equation}
be a finite collection of error operators, and define
\begin{equation}
    f_{ij}^{(N)}(m,m')
    :=
    \langle C_m|E_i^\dagger E_j|C_{m'}\rangle.
    \label{eq:fij-moving-band}
\end{equation}
Suppose that for every pair \(i,j\), there exist constants
\(\alpha_{ij}(c)\in\mathbb C\), \(C_{ij}\geq0\), and \(L_{ij}\geq0\),
independent of \(N\), such that
\begin{align}
    \left|
    f_{ij}^{(N)}(m,m')
    \right|
    &\leq
    \frac{C_{ij}}{\sqrt{N}},
    \qquad
    m\neq m',
    \label{eq:offdiag-moving-band}
    \\
    \left|
    f_{ij}^{(N)}
    \bigl(\bar m_N(c),\bar m_N(c)\bigr)
    -
    \alpha_{ij}(c)
    \right|
    &\leq
    \frac{C_{ij}}{\sqrt{N}},
    \label{eq:center-moving-band}
    \\
    \left|
    f_{ij}^{(N)}(m,m)
    -
    f_{ij}^{(N)}(n,n)
    \right|
    &\leq
    \frac{L_{ij}}{N}|m-n|,
    \nonumber \\
&     m,n\in B_N(c,K).
    \label{eq:Lipschitz-moving-band}
\end{align}

For any \(m\in B_N(c,K)\), it follows that
\begin{align}
    \left|
    f_{ij}^{(N)}(m,m)-\alpha_{ij}(c)
    \right|
    &\leq
    \left|
    f_{ij}^{(N)}(m,m)
    -
    f_{ij}^{(N)}
    \bigl(\bar m_N(c),\bar m_N(c)\bigr)
    \right|
    \nonumber\\
    &\quad+
    \left|
    f_{ij}^{(N)}
    \bigl(\bar m_N(c),\bar m_N(c)\bigr)
    -
    \alpha_{ij}(c)
    \right|
    \nonumber\\
    &\leq
    \frac{K L_{ij}+C_{ij}}{\sqrt{N}}.
    \label{eq:diag-moving-bound}
\end{align}
Hence,
\begin{equation}
    f_{ij}^{(N)}(m,m')
    =
    \alpha_{ij}(c)\delta_{mm'}
    +
    O\left(N^{-1/2}\right)
    \label{eq:KL-moving-band}
\end{equation}
uniformly for \(m,m'\in B_N(c,K)\).

Let
\begin{equation}
    \alpha(c)
    :=
    [\alpha_{ij}(c)]
\end{equation}
and let \(U(c)\) be a unitary matrix satisfying
\begin{equation}
    U(c)\alpha(c)U^\dagger(c)
    =
    \operatorname{diag}
    \left(
    \lambda_1(c),\ldots,\lambda_r(c)
    \right).
    \label{eq:alpha-c-diagonalization}
\end{equation}
The corresponding error operators are
\begin{equation}
    F_k^{(c)}
    :=
    \sum_{i=1}^{r}U_{ki}(c)E_i.
    \label{eq:rotated-errors-c}
\end{equation}
Then
\begin{equation}
    \langle C_m|
    F_k^{(c)\dagger}F_\ell^{(c)}
    |C_{m'}\rangle
    =
    \lambda_k(c)
    \delta_{k\ell}\delta_{mm'}
    +
    O\left(N^{-1/2}\right)
    \label{eq:approx-KL-c}
\end{equation}
uniformly throughout \(B_N(c,K)\).

Therefore, for each fixed value of \(c\), the approximate
Knill--Laflamme conditions become asymptotically exact in an
\(O(\sqrt{N})\) band around \(m=cN\).

\subsection{Constancy of the overlap matrix and its eigenvalues}

Let
\begin{equation}
    w_N(m)
    :=
    \left[
    f_{ij}^{(N)}(m,m)
    \right]_{i,j=1}^{r}.
\end{equation}
Equations~\eqref{eq:center-moving-band} and
\eqref{eq:Lipschitz-moving-band} imply
\begin{equation}
    \left\|
    w_N(m)-\alpha(c)
    \right\|
    \leq
    \frac{A(c,K)}{\sqrt{N}},
    \qquad
    m\in B_N(c,K),
    \label{eq:w-matrix-constant-c}
\end{equation}
where \(A(c,K)\) is independent of \(N\).  In particular, for any
\(m,n\in B_N(c,K)\),
\begin{equation}
    \left\|
    w_N(m)-w_N(n)
    \right\|
    \leq
    \frac{2A(c,K)}{\sqrt{N}}.
    \label{eq:w-matrix-pairwise}
\end{equation}
Thus, the full Knill--Laflamme matrix, rather than only its
eigenvectors, is approximately constant throughout the band.

Let \(\lambda_a^{(N)}(m)\) denote the eigenvalues of \(w_N(m)\).
By Weyl's eigenvalue perturbation inequality,
\begin{equation}
    \left|
    \lambda_a^{(N)}(m)-\lambda_a(c)
    \right|
    \leq
    \left\|
    w_N(m)-\alpha(c)
    \right\|
    \leq
    \frac{A(c,K)}{\sqrt{N}}.
    \label{eq:eigenvalue-constant-c}
\end{equation}
Consequently, both the overlap matrix and all its eigenvalues are
constant over the \(O(\sqrt N)\) band up to
\(O(N^{-1/2})\).

If an eigenvalue of \(\alpha(c)\) is separated from the remaining
spectrum by a nonzero gap \(g(c)\), the corresponding eigenvector
also changes by at most
\begin{equation}
    O\left(
    \frac{1}{g(c)\sqrt{N}}
    \right).
    \label{eq:eigenvector-constant-c}
\end{equation}
If degeneracies are present, the corresponding invariant
eigenspaces, rather than individual eigenvectors, are the
well-defined approximately constant objects.

\subsection{Single-qubit depolarizing channel}

Consider the single-qubit depolarizing channel
\begin{equation}
    \mathcal D_p(\rho)
    =
    \sum_{j=0}^{3}
    E_j^{(n)}\rho E_j^{(n)\dagger},
\end{equation}
with
\begin{equation}
    E_0^{(n)}
    =
    \sqrt{1-p}\,I,
    \qquad
    E_j^{(n)}
    =
    \sqrt{\frac{p}{3}}\,
    \sigma_n^j,
    \qquad
    j\in\{x,y,z\}.
\end{equation}
Define
\begin{equation}
    q
    :=
    \sqrt{\frac{p(1-p)}{3}},
\end{equation}
and
\begin{equation}
    D_m^x
    :=
    \frac{1}{2N}
    \sqrt{(N+m)(N-m-2)},
    \qquad
    D_m^z
    :=
    \frac{m}{N}.
    \label{eq:DxDz}
\end{equation}
The effective Knill--Laflamme matrix in the Pauli basis
\(\{I,\sigma^x,\sigma^y,\sigma^z\}\) is
\begin{equation}
    w_N(m)
    =
    \begin{pmatrix}
    1-p
        & qD_m^x
        & 0
        & qD_m^z
    \\
    qD_m^x
        & p/3
        & i(p/3)D_m^z
        & 0
    \\
    0
        & -i(p/3)D_m^z
        & p/3
        & i(p/3)D_m^x
    \\
    qD_m^z
        & 0
        & -i(p/3)D_m^x
        & p/3
    \end{pmatrix}.
    \label{eq:depolarizing-w-m}
\end{equation}

Let
\begin{equation}
    m=cN+\delta m,
    \qquad
    |\delta m|\leq K\sqrt{N}.
\end{equation}
Then
\begin{equation}
    D_m^z
    =
    c+\frac{\delta m}{N}
    =
    c+O\left(N^{-1/2}\right).
    \label{eq:Dz-c-expansion}
\end{equation}
Furthermore,
\begin{align}
    D_m^x
    &=
    \frac{1}{2}
    \sqrt{
    \left(1+\frac{m}{N}\right)
    \left(1-\frac{m}{N}-\frac{2}{N}\right)
    }
    \nonumber\\
    &=
    d_x(c)
    +
    O\left(N^{-1/2}\right),
    \label{eq:Dx-c-expansion}
\end{align}
where
\begin{equation}
    d_x(c)
    :=
    \frac{1}{2}\sqrt{1-c^2}.
    \label{eq:dx-c}
\end{equation}
The limiting overlap matrix is therefore
\begin{equation}
    \alpha(c)
    =
    \begin{pmatrix}
    1-p
        & qd_x(c)
        & 0
        & qc
    \\
    qd_x(c)
        & p/3
        & i(p/3)c
        & 0
    \\
    0
        & -i(p/3)c
        & p/3
        & i(p/3)d_x(c)
    \\
    qc
        & 0
        & -i(p/3)d_x(c)
        & p/3
    \end{pmatrix}.
    \label{eq:alpha-c-depolarizing}
\end{equation}
Using Eqs.~\eqref{eq:Dz-c-expansion} and
\eqref{eq:Dx-c-expansion}, we obtain
\begin{equation}
    \left\|
    w_N(m)-\alpha(c)
    \right\|
    =
    O\left(N^{-1/2}\right)
    \label{eq:depolarizing-w-convergence}
\end{equation}
uniformly for \(m\in B_N(c,K)\).  Hence the matrix, its
eigenvalues, and its invariant eigenspaces become independent of
\(m\) throughout the band.

This proves that the single-qubit depolarizing channel satisfies
the approximate Knill--Laflamme conditions in any
\(O(\sqrt N)\) band centered at a fixed value \(m/N=c\), and not
only in the equatorial band \(c=0\).
{
The band $B_N(c,K)$ remains in the physical range $m\in[-N/2,N/2]$ only if$|c|\leq \frac12-\frac{K}{\sqrt N}$,
which excludes an $O(N^{-1/4})$ neighborhood of each pole. This comes from using a fixed band $K\sqrt N$: since a coherent state's natural  deviation
is $\sqrt N\sin\theta$, a $\theta$-adapted radius remains physical up to an $O(N^{-1/4})$ neighborhood. Moreover, the exact single-error calculation in Appendix \ref{app:deformation}, gives an $O(N^{-1})$ decoded state deviation throughout this region, vanishing exactly at the poles. Thus, the restriction comes from the chosen band construction rather than from the code's actual performance near the poles.}
\subsection{Recovery independent of the unknown value of \(c\)}

The limiting matrix \(\alpha(c)\) generally depends on \(c\).
Therefore, the recovery obtained by exactly diagonalizing
\(\alpha(c)\) is also generally \(c\)-dependent.  If \(c\) is
unknown, one may instead use the \(c\)-independent reference matrix
\begin{equation}
    \alpha_{\mathrm{univ}}
    :=
    \operatorname{diag}
    \left(
    1-p,\frac{p}{3},\frac{p}{3},\frac{p}{3}
    \right).
    \label{eq:universal-alpha}
\end{equation}
The difference between the exact limiting matrix and this
universal matrix contains only off-diagonal terms.  Its norm is
\begin{equation}
    \left\|
    \alpha(c)-\alpha_{\mathrm{univ}}
    \right\|_{\mathrm F}^2
    =
    2\left[d_x(c)^2+c^2\right]
    \left(
    q^2+\frac{p^2}{9}
    \right).
    \label{eq:universal-alpha-bound1}
\end{equation}
Since the physical range satisfies \(|c|\leq1/2\),
\begin{equation}
    d_x(c)^2+c^2
    =
    \frac{1-c^2}{4}+c^2
    \leq
    \frac{7}{16}.
\end{equation}
It follows that
\begin{equation}
    \left\|
    \alpha(c)-\alpha_{\mathrm{univ}}
    \right\|
    \leq
    \sqrt{
    \frac{7}{8}
    \left(
    \frac{p(1-p)}{3}
    +
    \frac{p^2}{9}
    \right)
    }
    =
    O(\sqrt p).
    \label{eq:universal-alpha-bound2}
\end{equation}
Combining this with
Eq.~\eqref{eq:depolarizing-w-convergence} gives
\begin{equation}
    \left\|
    w_N(m)-\alpha_{\mathrm{univ}}
    \right\|
    \leq
    O\left(N^{-1/2}\right)
    +
    O(\sqrt p),
    \label{eq:universal-KL-bound}
\end{equation}
uniformly for all fixed values of \(c\).

Therefore, in the small-\(p\) regime, the original Pauli error
basis approximately diagonalizes the Knill--Laflamme matrix for
every value of \(m/N\).  A single recovery operation, independent
of the unknown value of \(m\) or \(c\), may consequently be used.

We emphasize the distinction between the two limits.  For a fixed
\(c\), the \(m\)-dependence within an \(O(\sqrt N)\) band vanishes
as \(O(N^{-1/2})\).  However, if one requires a single recovery
that works uniformly for all possible \(c\), the remaining
approximation error is controlled by \(p\):
\begin{equation}
    \epsilon_{\mathrm{univ}}
    =
    O\left(N^{-1/2}\right)+O(\sqrt p).
\end{equation}
Thus, for fixed small \(p\), the recovery is universally
approximate, but its error does not vanish completely as
\(N\rightarrow\infty\).  An asymptotically exact universal
recovery requires \(p\rightarrow0\) in addition to
\(N\rightarrow\infty\).



\subsection{Bandwidth--accuracy tradeoff}

More generally, consider a band
\begin{equation}
    B_N^{(\beta)}(c,K)
    :=
    \left\{
    m:
    |m-cN|\leq K N^\beta
    \right\},
    \qquad
    0<\beta<1.
\end{equation}
If the diagonal overlap functions are Lipschitz in \(m/N\), then
\begin{equation}
    w_N(m)-w_N(cN)
    =
    O\left(N^{\beta-1}\right).
\end{equation}
The choice \(\beta=1/2\), corresponding to the natural width of a
spin coherent state, gives
\begin{equation}
    w_N(m)-w_N(cN)
    =
    O\left(N^{-1/2}\right).
\end{equation}
Narrower bands give faster convergence, while wider sublinear
bands remain approximately correctable with a slower convergence
rate.
}
}

\section{Error threshold simulations}
\label{app:errsim}

The error threshold simulations for the ideal measurement case are performed in the density matrix formulation following the steps 0) to 4) in Sec. \ref{sec:threshold}. Here we describe how measurement imperfections in the form of initialization errors, measurement errors, and two-qubit errors are taken into account.  

\subsubsection{Measurement errors} 

We first describe measurement errors.  In our case, this amounts to physically performing a projection $ P_{sl} $ but having readout take values that are $ (s', l') $ with some probability.  This affects the correction operation since the unitary $ U_{s' l'} $ is applied instead, which is potentially the incorrect correction operation.  

This can be taken into account by defining a completely positive map
\begin{align}
    {\cal C}_q  (\rho) = \sum_{q'=1}^{q_\text{max} } C_{q q'} \rho C_{q q'} ^\dagger ,
\end{align}
where the index $ q $ enumerates the various $ (s,l) $ spaces in descending order, e.g. $ (2,1), (1,1), (1,2), (1,3), (0,1), (0,2) $ for $ N = 4$.   The Kraus operator $ C_{q q'} $ corresponds to the physical measurement outcome $ q $ but applying the correction operation for the outcome $ q' $.  We therefore take it as 
\begin{align}
   C_{q q'} = \sqrt{p_c(q,q')} U_{q'}
   \label{krausmeaserr}
\end{align}
where $ p_c(q,q') $ is the probability distribution for the physical measurement $ q $ but obtaining the readout $ q' $.  As such it satisfies $ \sum_{q'} p_c(q,q') =1 $.  In our simulations we take the probability distribution to be
\begin{align}
   p_c(q,q') =  \left\{
   \begin{array}{cc}
   1-p_\text{m} & \text{if } q'=q \\
   \frac{p_\text{m}}{2} & \text{if } |q'-q |=1 \\
   0 & \text{otherwise}
   \end{array} \right.    .
\end{align}
which amounts to a readout error with probability $ p_\text{m} $ which differs to the true projection by one.  At the boundaries $ q = 1, q_{\max} $, if the readout $ q' $ is out of bounds, it is interpreted as the closest outcome that is within bounds. For example, for $ q = 1 $ we take
\begin{align}
   p_c(1,q') =  \left\{
   \begin{array}{cc}
   1- \frac{p_\text{m}}{2} & \text{if } q'=1 \\
   \frac{p_\text{m}}{2} & \text{if } q'=2 \\
   0 & \text{otherwise}
   \end{array} \right.    .
\end{align}

Using this formalism, the syndrome measurement with measurement errors, with the correction step can be written as
\begin{align}
\rho \rightarrow  \sum_{q'=1}^{q_\text{max} }  \sum_{q=1}^{q_\text{max} } C_{q q'} P_q \rho P_q^\dagger C_{q q'}^\dagger ,
\label{genprojcorrect}
\end{align}
which generalizes (\ref{projcorrect}).  When including measurement errors, we replace (\ref{projcorrect}) with (\ref{genprojcorrect}) in Step 2, but otherwise the simulation proceeds in the same way.

\subsubsection{Initialization errors} 

Initialization errors can be handled using the same formalism as measurement errors.  As described in Ref. \cite{fowler2009high}, initialization error correspond to the ancilla qubits of the measurement circuit being intialized the incorrect value.  This again amounts to a false syndrome measurement readout. In our simulations, we use (\ref{genprojcorrect}) as the model of initialization error. The same Kraus operators (\ref{krausmeaserr}) are used but with the replacement $ p_{\text{m}} \rightarrow p_{\text{i}}  $, where $ p_{\text{i}} $ is the initialization error probability.




\end{document}